\documentclass[pre,twocolumn,longbibliography,10pt]{revtex4-1}

\usepackage{amssymb,amsmath}

\usepackage{graphicx}
\usepackage{xcolor}

\newcommand\beq{\begin{equation}}
\newcommand\eeq{\end{equation}}
\newcommand\beqa{\begin{eqnarray}}
\newcommand\eeqa{\end{eqnarray}}

\newcommand{\al}{\alpha}

\begin{document}
\title{Rheology of granular particles immersed in a molecular gas under uniform shear flow}
\author{Rub\'en G\'omez Gonz\'alez\footnote[1]{Electronic address: ruben@unex.es}}
\affiliation{Departamento de F\'{\i}sica,
Universidad de Extremadura, E-06006 Badajoz, Spain}
\author{Mois\'es Garc\'{\i}a Chamorro\footnote[1]{Electronic address: moises@unex.es}}
\affiliation{Departamento de F\'{\i}sica, Universidad de Extremadura, E-06006 Badajoz, Spain}
\author{Vicente Garz\'{o}\footnote[2]{Electronic address: vicenteg@unex.es;
URL: https://fisteor.cms.unex.es/investigadores/vicente-garzo-puertos/}}
\affiliation{Departamento de F\'{\i}sica and Instituto de Computaci\'on Cient\'{\i}fica Avanzada (ICCAEx), Universidad de Extremadura, E-06006 Badajoz, Spain}

 \begin{abstract}
 Non-Newtonian transport properties of a dilute gas of inelastic hard spheres immersed in a molecular gas are determined. We assume that the granular gas is sufficiently rarefied and hence, the state of the molecular gas is not disturbed by the presence of the solid particles. In this situation, one can treat the molecular gas as a bath (or thermostat) of elastic hard spheres at a given temperature. Moreover, in spite of the fact that the number density of grains is quite small, we take into account their inelastic collisions among themselves in its kinetic equation. The system (granular gas plus a bath of elastic hard spheres) is subjected to a simple (or uniform) shear flow (USF). In the low-density regime, the rheological properties of the granular gas are determined by solving the Boltzmann kinetic equation by means of Grad's moment method. These properties turn out to be highly nonlinear functions of the shear rate and the remaining parameters of the system. Our results show that the kinetic granular temperature and the non-Newtonian viscosity present a discontinuous shear thickening (DST) effect for sufficiently high values of the mass ratio $m/m_g$ ($m$ and $m_g$ being the mass of grains and gas particles, respectively). This effect becomes more pronounced as the mass ratio $m/m_g$ increases. In particular, in the Brownian limit ($m/m_g\to \infty$) the expressions of the non-Newtonian transport properties derived here are consistent with those previously obtained by considering a coarse-grained approach where the effect of gas phase on grains is through an effective force. Theoretical results are compared against computer simulations showing an excellent agreement.             
 \end{abstract}

 \maketitle

\section{Introduction}
\label{sec1}

Understanding the flow of solid particles immersed in an interstitial fluid is likely one of the most challenging open problems in granular flows. Among the different types of multiphase flows, the so-called particle-laden suspensions \cite{S20} (where dilute particles flow in a carrier fluid) constitutes one of the most studied problems in the granular literature. When collisions (which are generally assumed to be nearly instantaneous) play a relevant role in such suspensions, the kinetic theory conveniently adapted to account for the dissipative dynamics of grains can be considered as an useful tool to analyze the behavior of these sort of granular suspensions.     

However, due to the technical difficulties embodied in the description of two or more phases, a coarse-grained approach is usually adopted \cite{K90,G94,J00,FH17}. In this approach, the effect of the surrounding gas on solid particles is taken into account through an effective fluid-solid force. 
While in some works \cite{G94,ZW21,CBC23} the gas–solid force is simply proportional to the relative mean flow velocity between the solid and gas phases, in other works the force also incorporates a term proportional to the velocity particle (Stokes linear drag law) \cite{TK95,SMTK96,WZLH09,H13,ChVG15,SA17,ASG19,SA20} 
plus a Langevin-like stochastic term \cite{GTSH12}. While the Stokes drag force term accounts for the energy dissipated by grains due to their friction on the viscous gas, the stochastic term takes into account the energy transferred to grains due to their “interactions” with the background particles. When both terms (viscous drag force plus Langevin-like term) are considered, a Fokker–Planck term is incorporated in the corresponding kinetic equation.    

For small spatial gradients, the above suspension model \cite{GTSH12} has been solved by means of the Chapman--Enskog method \cite{CC70} to determine the Navier--Stokes transport coefficients in terms of the coefficient of restitution and the parameters of the system \cite{GTSH12,GChV13a,GChV13b,GG23}. The knowledge of the transport coefficients has allowed to assess the impact of the surrounding gas on the dynamic properties of grains. Beyond the Navier--Stokes regime, this sort of effective models have been also employed to obtain the non-Newtonian transport properties in (uniform) \emph{sheared} granular suspensions \cite{TK95,SMTK96,H13,SMMD13,KIB14,ChVG15,SA17,HTG17,HTG20,HT19,SA20,THSG20,THG23}. Theoretical and computational results have clearly shown the existence in sheared gas-solid flows of the so-called DST, namely, the non-Newtonian shear viscosity of the suspension drastically increases with increasing the shear rate.

Although the above suspension models (which are based on a Markovian form of the Langevin equation) yield good agreement with computer simulations \cite{CD13,CDF15,GFHY16,FLYH17}, a recent work \cite{PZ23} has questioned the use of the standard Langevin equation in sheared suspensions. In fact, the main conclusion of this paper \cite{PZ23} is that the usual Markovian Langevin equation widely employed in the granular literature is valid only in the limit of extremely weak shear rates compared to the effective vibrational temperature of the bath. For high-shear rates (where the DST effect appears), the starting point should be a (generalized) non-Markovian Langevin equation related with a nontrivial fluctuation-dissipation theorem \cite{PZ23}. It is important to remark that this assessment has been achieved from first principles and hence, there is no doubt on the robustness of this conclusion. Thus, to shed light on this controversy and justify the conditions under which the conventional Langevin equation in the theory of Brownian motion of sheared granular suspensions can be used, it would be convenient to propose a suspension model where the real collisions between grains and particles of the molecular gas were considered. A possibility would be to start from a set of two coupled kinetic equations for the one-particle velocity distribution functions of the solid and gas phases. However, the determination of transport properties of the solid particles from this suspension model would be in general a quite intricate task since in particular the different phases evolve over quite different spatial and temporal scales. 

To overcome the above technical difficulties, a suspension model inspired in a paper of Biben \emph{et al.} \cite{BMP02a} has been recently proposed \cite{GG22a}. In this model, one assumes first that the concentration of the granular particles (or ``granular gas'') is much smaller than that of the molecular gas and so, one can assume that the state of the latter is not perturbed by the presence of grains. This assumption can be justified in the case of particle-laden suspensions where \emph{dilute} particles are immersed in a carried fluid (for instance, fine aerosol particles in air). Under these conditions, the background (molecular) gas can be treated as a bath or \emph{thermostat} at a certain temperature $T_g$. Needless to say, this suspension model (granular particles immersed in a molecular gas) can be considered as the \emph{tracer} limit of a binary mixture; the grains of course are the defect species of the mixture. In addition, although the number density of grains is very small, one has to take into account not only the \emph{elastic} collisions between solid and gas particles in the kinetic equation of grains, but also the \emph{inelastic} collisions among grains themselves. Thus, in contrast to the coarse-grained approaches \cite{K90,G94,J00,GTSH12,FH16}, two new input parameters are introduced in this suspension model: the diameter $\sigma/\sigma_g$ and mass $m/m_g$ ratios. Here, $\sigma_g$ and $m_g$ are the diameter and mass of the particles of the molecular gas, respectively, while $\sigma$ and $m$ are the diameter and mass of the solid particles, respectively. In fact, as shown in Ref.\ \cite{GG22a}, the expressions of the Navier--Stokes transport coefficients derived from this collisional model reduce to those previously obtained from the coarse-grained approach \cite{GGG19a} when the grains are much heavier than the particles of the molecular gas (Brownian limit $m/m_g\to \infty$).

The goal of this paper is to determine the non-Newtonian transport properties of a granular suspension under uniform shear flow (USF). Here, in contrast to previous attempts \cite{TK95,SMTK96,H13,SMMD13,KIB14,ChVG15,SA17,HTG17,HT19,SA20,THSG20,THG23} carried out by starting from the Stokes and/or Langevin suspension models, we will start from the collisional model proposed in Ref.\ \cite{GG22a} where the collisions between the solid and molecular gas particles are taken into account. In the case of a binary mixture constituted by a granular gas thermostated by a molecular gas of elastic hard spheres, the USF state is characterized by constant number densities for solid and gas particles, a uniform temperature, and a (common) linear velocity profile $U_{g,x}=U_x=ay$, where $a$ is the constant shear rate. Here, $\mathbf{U}_g$ and $\mathbf{U}$ are the mean flow velocities of the molecular and granular gases, respectively. 

In the low-density regime, the distribution function $f_g(\mathbf{r}, \mathbf{v};t)$ of the molecular gas verifies the nonlinear Boltzmann equation for hard spheres. Since we are interested in a steady state, an external thermostat force must be introduced to compensate for the viscous heating effect produced by the shear field in the molecular gas. Here,   
as usual in non-equilibrium molecular dynamics simulations \cite{EM90}, we introduce a ``drag'' force  of the form $\mathbf{F}_g=-m_g \xi \mathbf{V}$, where $m_g$ is the mass of a particle of the molecular gas and $\mathbf{V}=\mathbf{v}-\mathbf{U}$ is the peculiar velocity. This sort of thermostat is usually referred to as the Gaussian thermostat. The quantity $\xi$ is a function of the shear rate adjusted as to keep the temperature $T_g$ constant. It is important to remark that only in the especial case of Maxwell molecules \cite{GS03}, there is an exact equivalence between the results derived with and without the Gaussian thermostat \cite{DSBR86}. Regarding the distribution function $f(\mathbf{r}, \mathbf{v};t)$ of the granular gas, its time evolution involves the Boltzmann $J[f,f]$ and Boltzmann--Lorentz $J_g[f,f_g]$ collision operators. While the first nonlinear operator $J[f,f]$ takes into account the rate of change of $f$ due to the inelastic collisions between grains themselves, the linear operator $J_g[f,f_g]$ accounts for the rate of change of $f$ due to the elastic collisions between particles of the molecular gas and grains. 

As in previous works \cite{HTG17,HTG20,HT19,THSG20} on sheared granular suspensions, the rheological properties of both the granular and molecular gases are approximately obtained by means of Grad's moment method \cite{G49}. In spite of the simplicity of Grad's method, it is important to remark that the rheological properties obtained from this method agrees in general quite well with computer simulations \cite{HTG17,HT19,THSG20,THG23} in the Brownian limit ($m/m_g\to \infty$), even for high values of the shear rate. Therefore, the conventional Grad's moment method can be still considered as a powerful method to
describe the rheology of gas-solid suspensions. In particular, theory and simulations have clearly shown the discontinuous transition of the kinetic temperature and nonlinear shear viscosity for dilute suspensions \cite{HT19,GGG19}. Moreover, as the density increases, there is a transition from the DST (observed in dilute gases) to the continuous shear thickening (CST) for denser systems \cite{HTG17,THSG20}.

The real contribution of the present paper to the understanding of the rheology of granular suspensions may be considered two-fold. First, as mentioned before and to the best of our knowledge, our results provide for the first time
the explicit dependence of the rheological properties on the mass and diameter ratios, apart from their dependence on the remaining parameters of the system (shear rate, coefficient of restitution for inelastic grain-grain collisions, and background temperature). 
These results clearly show how the DST effect increases with increasing the mass ratio $m/m_g$ (or, analogously, the diameter ratio $\sigma/\sigma_g$). As a second and relevant contribution,
we show how the shear-rate dependence of the rheological properties tend to those obtained by using a coarse-grained approach \cite{HT19,GGG19} as the mass ratio $m/m_g$ becomes more and more larger. Thus, our results (which apply for arbitrary values of the mass ratio) justify the use of the white assumption in the coarse-grained approach (Markovian Langevin equation) when the mass the granular particles is much heavier than that of the particles of the molecular gas (bath). This achievement can be also understood from the first principles results derived in Ref.\ \cite{PZ23}. Thus, if you look at Eq.\ (11) of this paper \cite{PZ23}, the shear term ($\dot{\gamma}q_{iy}p_{ix} \hat{\mathbf{x}}\hat{\mathbf{y}}$) in the bath Hamiltonian $\mathcal{H}_B$ is proportional to the mass of the bath particles 
When this term becomes negligible in the coupled particle-bath dynamics, the fluctuation-dissipation theorem remains unaltered by shear flow, as the bath's response to the shear rate becomes insignificant compared to that of the grains. This result makes the use of the  white noise assumption justified, even for large values of the shear rate. It is quite apparent that our results can be considered as a quantitative way to a posteriori justification of the use of the conventional Langevin equation when $m/m_g\to \infty$ as the starting point for studying the rheology of the granular suspension. This conclusion (not considered in previous works on gas-solid flows) represents one of the most significant added values of the paper and can likely justify by itself the need of the present work.

The plan of the paper is as follows. In Sec.\ \ref{sec2}, we present the Boltzmann kinetic equation for a granular gas in contact with a bath of elastic hard spheres. The balance equations for the densities of mass, momentum, and energy of the granular and molecular gases are also displayed. Section \ref{sec3} deals with the rheological properties of the molecular gas. As expected, a shear thinning effect for the nonlinear shear viscosity is observed from both Grad's solution and DSMC results. The rheological properties of the granular gas are determined in Sec.\ \ref{sec4}. In dimensionless form, they are given in terms of the mass ratio $m/m_g$, the diameter ratio $\sigma/\sigma_g$, the coefficient of restitution $\al$, the (reduced) bath temperature $T_g^*$, and the (reduced) shear rate $a^*=a/\gamma$. Here, $\gamma$ is the drift coefficient characterizing the friction of solid particles on the viscous gas. The Brownian limiting case ($m/m_g\to \infty$) is considered in Sec.\ \ref{sec5} where we show the consistency of the present results with those derived before \cite{HTG17,HT19,GGG19,THSG20} by employing a coarse-grained approach. The shear-rate dependence of the rheological properties of the granular gas is illustrated in Sec.\ \ref{sec6}. Here, we relate the diameter and mass ratios, fix the background temperature and consider different values of both the coefficient of restitution and the mass ratio. Our results highlight the existence of a DST effect as the mass ratio $m/m_g$ increases; the temperature ratio $\theta=T/T_g$ and the non-Newtonian shear viscosity $\eta^*$ discontinuosly increases/decreases (at a certain value of $a^*)$ as the (reduced) shear rate $a^*$ gradually increases/decreases. This effect is more pronounced as $m/m_g$ becomes more large. Theoretical results obtained from Grad's method are compared with computer simulations in the Brownian limit ($m/m_g\to \infty$) and for small mass ratios showing an excellent agreement between both approaches. We close the paper in Sec.\ \ref{sec7} with a brief summary and some concluding remarks on the results reported here.

\section{Granular gas thermostated by a bath of elastic hard spheres. Boltzmann kinetic theory}
\label{sec2}

Let us consider a gas of inelastic hard disks ($d=2$) or spheres ($d=3$) of mass $m$ and diameter $\sigma$. Inelasticity of collisions is characterized through a constant (positive) coefficient of normal restitution $\al\leq 1$. We assume that the granular gas is immersed in gas of
elastic hard disks or spheres (molecular gas) of mass $m_g$ and diameter $\sigma_g$. Collisions between granular particles and particles of the molecular gas are elastic. As mentioned in Sec.\ \ref{sec1}, we want to consider a situation where the number density of the granular gas $n$ is much smaller than that of the molecular gas $n_g$. In this case, the granular gas is sufficiently rarefied so that the state of the molecular gas is not affected by the presence of the solid (grains) particles. This means that the molecular gas may be considered as a thermostat at the temperature $T_g$. In addition, although the concentration of granular particles is small, the collisions among grains themselves are also accounted for in the corresponding kinetic equation for the one-particle velocity distribution function $f(\mathbf{r}, \mathbf{v}; t)$ of solid particles.

\subsection{Molecular gas}

Under the above conditions, in the low-density regime, the kinetic equation of the one-particle velocity distribution function $f_g(\mathbf{r}, \mathbf{v}; t)$ of the molecular gas is the Boltzmann kinetic equation \cite{C88,GS03}:
\beq
\label{2.1}
\frac{\partial f_g}{\partial t}+\mathbf{v}\cdot \nabla f_g+\frac{\partial}{\partial \mathbf{v}}\cdot \left(\frac{\mathbf{F}_g}{m_g}f_g\right)=J_g[f_g,f_g],
\eeq
where the Boltzmann collision operator $J_g[f_g,f_g]$ is 
\beqa
\label{2.2}
J_g[\mathbf{v}_1|f_g,f_g]&=&\sigma_g^{d-1}\int d\mathbf{v}_{2}\int d\widehat{\boldsymbol {\sigma
}}\Theta (\widehat{{\boldsymbol {\sigma }}}_g \cdot {\mathbf g}_{12}) (\widehat{\boldsymbol {\sigma }}\cdot {\mathbf g}_{12})
\nonumber\\
& & \times \left[f_g({\mathbf v}_{1}'')f_g({\mathbf v}_{2}'')-f_g({\mathbf v}_{1})f_g({\mathbf v}_{2})\right].
\eeqa
In Eq.\ \eqref{2.2}, $\mathbf{g}_{12}=\mathbf{v}_1-\mathbf{v}_2$ is the relative velocity, $\widehat{\boldsymbol {\sigma }}$ is a unit vector along the line of centers of the two spheres at contact, $\Theta$ is the Heaviside step function, and the double primes denote pre-collisional velocities. The relationship between pre-collisional $(\mathbf{v}_1'', \mathbf{v}_2'')$ and post-collisional $(\mathbf{v}_1, \mathbf{v}_2)$ velocities is 
\beq
\label{2.2.1}
\mathbf{v}_{1,2}''=\mathbf{v}_{1,2}\mp(\widehat{{\boldsymbol {\sigma }}} \cdot {\mathbf g}_{12})\widehat{\boldsymbol {\sigma}}.
\eeq
Moreover, we have assumed in Eq.\ \eqref{2.1} that the particles of the molecular gas can be subjected to the action of the (nonconservative) external force $\mathbf{F}_g$. 

The relevant hydrodynamic fields of the molecular gas are the number density $n_g(\mathbf{r}; t)$, the mean flow velocity $\mathbf{U}_g(\mathbf{r}; t)$, and the temperature $T_g(\mathbf{r}; t)$. These fields are defined in terms of the velocity distribution function $f_g$ as
\beq
\label{2.2a}
\left\{n_g, n_g \mathbf{U}_g, d n_g T_g \right\}=\int d\mathbf{v}\left\{1, \mathbf{v}, m_g V_g^2\right\}f_g(\mathbf{v}),
\eeq
where $\mathbf{V}_g=\mathbf{v}-\mathbf{U}_g$ is the peculiar velocity of the molecular gas. Note that in Eq.\ \eqref{2.2a} the Boltzmann constant $k_\text{B}=1$. We will take this value throughout the paper for the sake of simplicity. The balance equations for the hydrodynamic fields $n_g$, $\mathbf{U}_g$, and $T_g$ are \cite{GS03}
\beq
\label{2.3}
D_t^g n_g+n_g \nabla \cdot \mathbf{U}_g=0,
\eeq
\beq
\label{2.4}
\rho_g D_t^g \mathbf{U}_g+\nabla \cdot \mathsf{P}_g=\boldsymbol{\mathcal{F}}_{g}^U,
\eeq
\beq
\label{2.5}
\frac{d}{2}n_g D_t^g T_g+\nabla \cdot \mathbf{q}_g+\mathsf{P}_g:\nabla \mathbf{U}_g={\mathcal{F}}_{g}^T,
\eeq
where $D_t^g\equiv\partial_t+\mathbf{U}_g\cdot \nabla$ is the material time derivative for the molecular gas, $\rho_g=m_gn_g$ is the mass density of molecular gas,
\beq
\label{2.6}
\mathsf{P}_g=m_g \int d\mathbf{v}\; \mathbf{V}_g\mathbf{V}_g f_g(\mathbf{v})
\eeq 
is the pressure tensor,   
\beq
\label{2.7}
\mathbf{q}_g=\frac{m_g}{2}\int d\mathbf{v}\; V_g^2 \mathbf{V}_g f_g(\mathbf{v})
\eeq
is the heat flux, 
\beq
\label{2.8}
\boldsymbol{\mathcal{F}}_{g}^U=\int d\mathbf{v}\; \mathbf{F}_g f_g(\mathbf{v}), \quad 
{\mathcal{F}}_{g}^T=\int d\mathbf{v}\; \mathbf{V}_g\cdot \mathbf{F}_g f_g(\mathbf{v})
\eeq
are the production terms of momentum and energy, respectively, due to the external force $\mathbf{F}_g$.

\subsection{Granular gas}

Since the particles of the granular gas collide among themselves and with particles of the molecular gas, its distribution function $f(\mathbf{r}, \mathbf{v}; t)$ in the low-density regime verifies the kinetic equation 
\beq
\label{2.9}
\frac{\partial f}{\partial t}+\mathbf{v}\cdot \nabla f+\frac{\partial}{\partial \mathbf{v}}\cdot \left(\frac{\mathbf{F}}{m}f\right)=J[f,f]+J_\text{BL}[f,f_g],
\eeq
where $\mathbf{F}$ is the external force acting on solid particles. As said in Sec.\ \ref{sec1}, the Boltzmann collision operator $J[f,f]$ gives the rate of change of the distribution $f$ due to binary inelastic collisions between granular particles while the Boltzmann-Lorentz collision operator $J_\text{BL}[f,f_g]$ accounts for the rate of change of the distribution $f$ due to elastic collisions between granular and molecular gas particles. The explicit form of the nonlinear Boltzmann collision operator $J[f,f]$ is \cite{G19}
\beqa
\label{2.10}
J[\mathbf{v}_1|f,f]&=&\sigma^{d-1}\int d\mathbf{v}_{2}\int d\widehat{\boldsymbol {\sigma
}}\Theta (\widehat{{\boldsymbol {\sigma }}} \cdot {\mathbf g}_{12}) (\widehat{\boldsymbol {\sigma }}\cdot {\mathbf g}_{12})
\nonumber\\
& & \times 
\left[\al^{-2}f({\mathbf v}_{1}'')f({\mathbf v}_{2}'')-f({\mathbf v}_{1})f({\mathbf v}_{2})\right],
\eeqa
where in Eq.\ \eqref{2.10} the relationship between pre-collisional $(\mathbf{v}_1'', \mathbf{v}_2'')$ and post-collisional $(\mathbf{v}_1, \mathbf{v}_2)$ velocities is
\beq
\label{2.11}
\mathbf{v}_{1,2}''=\mathbf{v}_{1,2}\mp\frac{1+\al}{2\al}(\widehat{{\boldsymbol {\sigma }}} \cdot {\mathbf g}_{12})\widehat{\boldsymbol {\sigma }}.
\eeq
The form of the linear Boltzmann-Lorentz collision operator $J_\text{BL}[f,f_g]$ is \cite[]{G19,RL77}
\beqa
\label{2.12}
J_\text{BL}[\mathbf{v}_1|f,f_g]&=&\overline{\sigma}^{d-1}\int d\mathbf{v}_{2}\int d\widehat{\boldsymbol {\sigma
}}\Theta (\widehat{{\boldsymbol {\sigma }}} \cdot {\mathbf g}_{12}) (\widehat{\boldsymbol {\sigma }}\cdot {\mathbf g}_{12})
\nonumber\\
& & \times \left[f({\mathbf v}_{1}'')f_g({\mathbf v}_{2}'')-f({\mathbf v}_{1})f_g({\mathbf v}_{2})\right],
\eeqa
where $\overline{\sigma}=(\sigma+\sigma_g)/2$ and in Eq.\ \eqref{2.12} the relationship between $(\mathbf{v}_1'',\mathbf{v}_2'')$ and $(\mathbf{v}_1,\mathbf{v}_2)$ is
\beq
\label{2.13}
\mathbf{v}_1''=\mathbf{v}_1-2 \mu_g(\widehat{{\boldsymbol {\sigma }}} \cdot {\mathbf g}_{12})\widehat{\boldsymbol {\sigma }}, \quad
\mathbf{v}_2''=\mathbf{v}_2+2 \mu (\widehat{{\boldsymbol {\sigma }}} \cdot {\mathbf g}_{12})\widehat{\boldsymbol {\sigma }}.
\eeq
Here, 
\beq
\label{2.14}
\mu_g=\frac{m_g}{m+m_g}, \quad  \mu=\frac{m}{m+m_g}.
\eeq

As in the case of the molecular gas, the relevant hydrodynamic fields of the granular gas are the number density 
\beq
\label{2.15}
n(\mathbf{r}; t)=\int d\mathbf{v}\; f(\mathbf{v}),
\eeq
the mean flow velocity
\beq
\label{2.16}
\mathbf{U}(\mathbf{r}; t)=\frac{1}{n(\mathbf{r}; t)}\int d\mathbf{v}\; \mathbf{v} f(\mathbf{v}),
\eeq
and the \emph{granular} temperature 
\beq
\label{2.17}
T(\mathbf{r}; t)=\frac{m}{d n(\mathbf{r}; t)}\int d\mathbf{v}\; V^2 f(\mathbf{v}),
\eeq
where $\mathbf{V}=\mathbf{v}-\mathbf{U}$ is the peculiar velocity of the granular gas. Note that in general the mean flow velocity $\mathbf{U}$ of solid particles can be different from the mean flow velocity $\mathbf{U}_g$ of particles of the molecular gas. In fact, the difference $\mathbf{U}-\mathbf{U}_g$ may induce a non-vanishing contribution to the heat flux \cite{GG22a}. 

The macroscopic balance equations for the granular gas are obtained by multiplying Eq.\ \eqref{2.9} by $\left\{1, \mathbf{v}, m V^2\right\}$ and integrating over velocity. After some algebra, one gets the result 
\beq
\label{2.18}
D_t n+n\nabla\cdot \mathbf{U}=0,
\eeq
\beq
\label{2.19}
\rho D_t\mathbf{U}+\nabla \cdot \mathsf{P}=\boldsymbol{\mathcal{F}}^U+\boldsymbol{\mathcal{F}}[f],
\eeq
\beq
\label{2.20}
\frac{d}{2}n D_t T+\nabla \cdot \mathbf{q}+\mathsf{P}:\nabla \mathbf{U}={\mathcal{F}}^T- \frac{d}{2} n T\left(\zeta+\zeta_g\right).
\eeq
Here, $D_t=\partial_t+\mathbf{U}\cdot \nabla$ is the material derivative of solid particles, $\rho=m n$ is the mass density of solid particles, and the pressure tensor $\mathsf{P}$ and the heat flux vector $\mathbf{q}$ are given, respectively, as
\beq
\label{2.21}
\mathsf{P}=\int d\mathbf{v}\; m \mathbf{V}\mathbf{V} f(\mathbf{v}),
\eeq
\beq
\label{2.22}
\mathbf{q}=\int d\mathbf{v}\; \frac{m}{2} V^2 \mathbf{V} f(\mathbf{v}).
\eeq
The production terms of momentum and energy due to the external force $\mathbf{F}$ are 
\beq
\label{2.23}
\boldsymbol{\mathcal{F}}^U=\int d\mathbf{v}\; \mathbf{F} f(\mathbf{v}), \quad
{\mathcal{F}}^T=\int d\mathbf{v}\; \mathbf{V}\cdot \mathbf{F} f(\mathbf{v}).
\eeq
Since the Boltzmann--Lorentz collision term $J_g[f,f_g]$ does not conserve momentum, then the production of momentum term $\boldsymbol{\mathcal{F}}[f]$ due to collisions between the granular and molecular particles is in general different from zero. It is defined as
\beq
\label{2.24}
\boldsymbol{\mathcal{F}}[f]=\int d\mathbf{v}\; m \mathbf{V}  J_\text{BL}[f,f_g].
\eeq
For small spatial gradients, the production term $\boldsymbol{\mathcal{F}}[f]$ vanishes when $\mathbf{U}=\mathbf{U}_g$ \cite{GG22a}. In addition, the partial production rates $\zeta$ and $\zeta_g$ are given, respectively, as
\beq
\label{2.25}
\zeta=-\frac{m}{d n T}\int d\mathbf{v}\; V^2\; J[\mathbf{v}|f,f],
\eeq
\beq
\label{2.26}
\zeta_g=-\frac{m}{d n T}\int d\mathbf{v}\; V^2\; J_\text{BL}[\mathbf{v}|f,f_g].
\eeq
The cooling rate $\zeta$ gives the rate of kinetic energy loss due to inelastic collisions between particles of the granular gas. It vanishes for elastic collisions. On the other hand, the term $\zeta_g$ gives the transfer of kinetic energy between the particles of the granular and molecular gases and it cancels out in the homogeneous state when both gases are at the same temperature ($T_g=T$) \cite{GG22a}.

It is quite apparent that the balance equations \eqref{2.3}--\eqref{2.5} and \eqref{2.18}--\eqref{2.20} for the molecular and granular gases, respectively, do not constitute a closed set of equations for the hydrodynamic fields unless one expresses the fluxes, the production momentum term, and the cooling rates in terms of those fields. Here, we will consider an specific non-equilibrium state: the USF state.

\section{Granular particles immersed in a molecular gas under USF}
\label{sec3}

We assume that the system (granular particles plus molecular gas) are subjected to USF. At a macroscopic level, this state is characterized by a linear profile of the $x$ component of the flow velocities along the $y$ axis, constant densities $n_g$ and $n$, and uniform temperatures $T_g$ and $T$:
\beq
\label{3.2}
n_g\equiv \text{const}, \quad n\equiv \text{const},
\eeq
\beq
\label{3.3}
\nabla T_g=\nabla T=0,
\eeq
\beq
\label{3.1}
U_{g,i}=U_i=a_{ij}r_j, \quad a_{ij}=a\delta_{ix}\delta_{jy},
\eeq
$a$ being the \emph{constant} shear rate. Since the only spatial gradient present in the system is the shear rate, the heat fluxes $\mathbf{q}_g$ and $\mathbf{q}$ vanish and the pressure tensors $\mathsf{P}_g$ and $\mathsf{P}$ are the relevant fluxes in the problem. They measure the transport of momentum across the system. In addition, the production of momentum term $\boldsymbol{\mathcal{F}}[f]$ vanishes in the USF problem since $\mathbf{U}=\mathbf{U}_g$.

On the other hand, according to Eq.\ \eqref{2.5}, it is quite apparent that in the absence of an external force the USF for the molecular gas is not a stationary state since the temperature increases with time due to viscous heating. Here, as mentioned in Sec.\ \ref{sec1}, a nonconservative external force of the form \cite{EM90}
\beq
\label{3.4}
\mathbf{F}_g=-m_g \xi \mathbf{V}
\eeq
is introduced to achieve a steady state. The quantity $\xi$ is chosen to be a function of the shear rate by the condition that the temperature $T_g$ reaches a constant value in the long time limit. Analogously, the granular gas is also subjected to this sort of Gaussian thermostat so that, $\mathbf{F}=-m \xi \mathbf{V}$. Note that the parameter $\xi$ is common for both species. The use of this kind of thermostat is quite usual in sheared molecular mixtures \cite{GS03}. Since $\mathbf{F}_g$ and $\mathbf{F}$ are both proportional to the peculiar velocity $\mathbf{V}$, then $\boldsymbol{\mathcal{F}}_g^U=\boldsymbol{\mathcal{F}}^U=\mathbf{0}$, and 
\beq
\label{3.4.1}
\frac{\mathcal{F}_g^T}{n_g T_g}=\frac{\mathcal{F}^T}{n T}=-d \xi.     
\eeq

\subsection{Rheology of the molecular gas}

As the state of the molecular gas is not disturbed by the presence of granular particles, it is more convenient to determine its rheological properties before considering the USF state for the solid particles. At a microscopic level, the USF state becomes spatially homogeneous when the velocities of the particles are referred to the Lagrangian frame moving with the flow velocity $\mathbf{U}$. In this new frame, the distribution function of the molecular gas adopts the form $f_g(\mathbf{r},\mathbf{v})=f_g(\mathbf{V})$ and the Boltzmann equation \eqref{2.1} in the presence of the external force \eqref{3.4} reads
\beq
\label{3.5}
-\frac{\partial}{\partial V_i}\left(a_{ij}V_{j}+\xi V_i\right)f_g({\bf V})=J_g[f_g,f_g].
\end{equation}

The relevant transport coefficients of the USF problem are obtained from the knowledge of the non-zero elements of the pressure tensor $\mathsf{P}_g$. These elements can be determined by multiplying the Boltzmann equation \eqref{3.5} by $m_g{\bf V}{\bf V}$ and integrating over ${\bf V}$. The result is
\begin{equation}
\label{3.6}
a_{ik}P_{g,kj}+a_{j k}{P}_{g,ki}+2\xi P_{g,ij}={A}_{g,ij},
\end{equation}
where we have introduced the collisional moment $\mathsf{A}_g$ as
\begin{equation}
\label{3.7}
{A}_{g,ij}=m_g\int d{\bf V}\; V_i V_j J_{g}[{\bf V}|f_g,f_g].
\end{equation}
From Eq.\ \eqref{3.6}, in particular, one gets the balance equation for the temperature  $T_g$:
\begin{equation}
\label{3.8}
aP_{g,xy}+d \xi p_g =0,
\end{equation}
where $p_g=n_gT_g$ is the hydrostatic pressure of the molecular gas.

The determination of the collisional moment $\mathsf{A}_g$ for arbitrary shear rate requires the knowledge of the velocity distribution function $f_g(\mathbf{V})$. This is a quite formidable task. However, one expects to get a good estimate of  this collisional moment by using Grad’s approximation \cite{G49}
\begin{equation}
\label{3.9}
f_g({\bf V})\to f_{g,\text{M}}({\bf V})\left(1+\frac{m_g}{2 T_g} \Pi_{g,ij}V_i V_j\right),
\end{equation}
where 
\begin{equation}
\label{3.10}
f_{g,\text{M}}({\bf V})=n_g \left(\frac{m_g}{2\pi T_g}\right)^{d/2}\exp\left(-\frac{m_g V^2}{2 T_g}\right)
\end{equation}
is the Maxwellian distribution of the molecular gas and 
\beq
\label{3.11}
\Pi_{g,ij}=\frac{P_{g,ij}}{p_g}-\delta_{ij}
\eeq 
is the traceless part of the (dimensionless) pressure tensor. With the trial distribution \eqref{3.9}, the collisional moment $\mathsf{A}_g$ can be explicitly evaluated. When one neglects nonlinear terms in the tensor $\boldsymbol{\Pi}_g$, the expression of $\mathsf{A}_g$ can be written as \cite{G02,SGD04}
\beq
\label{3.12}
A_{g,ij}=-\nu_g p_g{\Pi}_{g,ij},
\eeq
where the collision frequency $\nu_g$ is
\begin{equation}
\label{3.13}
\nu_{g}=\frac{8\pi^{(d-1)/2}}{(d+2)\Gamma(d/2)}n_g\sigma_g^{d-1}\sqrt{\frac{T_g}{m_g}}.
\end{equation}

\begin{figure}[ht]
\centering
\includegraphics[width=0.4\textwidth]{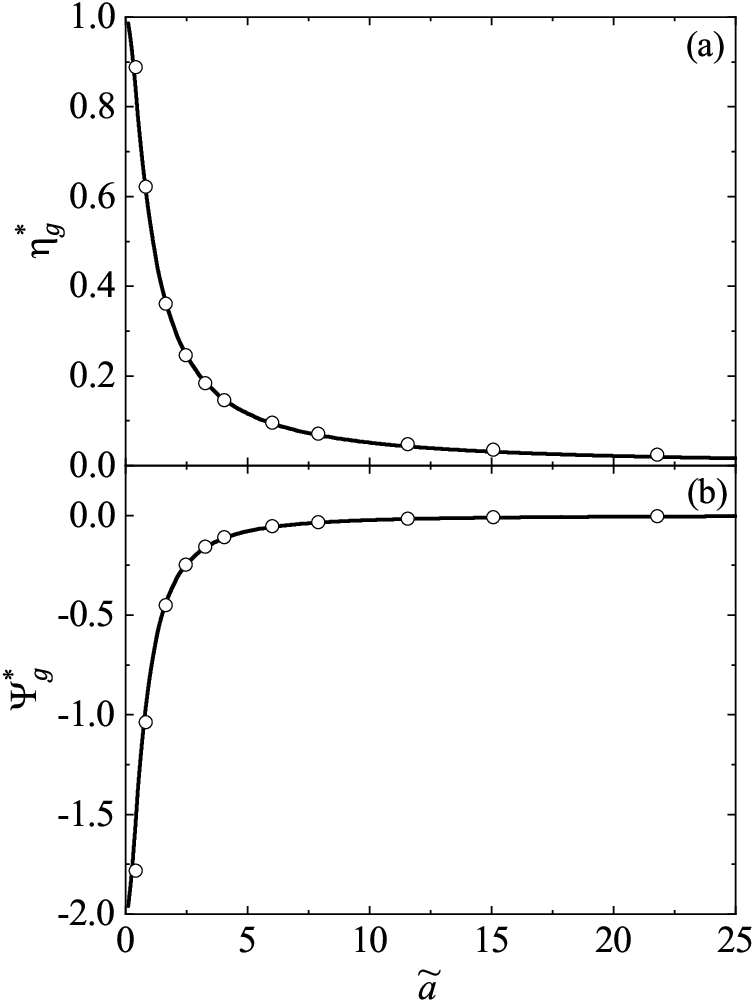}
\caption{Plot of the nonlinear shear viscosity $\eta_g^*$ [panel (a)] and first viscometric function $\Psi_g^*$ [panel (b)] for hard spheres ($d=3$) as functions of the (reduced) shear rate $\widetilde{a}=a/\nu_g$. Symbols refer to the DSMC results.}
\label{fig1}
\end{figure}

The non-zero elements of the pressure tensor $\mathsf{P}_g^*=\mathsf{P}_g/(n_g T_g)$ can be easily obtained from Eq.\ \eqref{3.6} when one takes into account Eq.\ \eqref{3.12}. They are given by
\begin{equation}
\label{3.14}
P_{g,xx}^*=\frac{1}{1+2\widetilde{\xi}}\left[1+\frac{2\widetilde{a}^2}{(1+2\widetilde{\xi})^2}\right],
\end{equation}
\begin{equation}
\label{3.15}
P_{g,yy}^*=P_{g,zz}^*=\frac{1}{1+2\widetilde{\xi}},
\end{equation}
\begin{equation}
\label{3.16}
P_{g,xy}^*=-\frac{\widetilde{a}}{(1+2\widetilde{\xi})^2},
\end{equation}
where we have introduced the dimensionless quantities 
\beq
\label{3.17}
\widetilde{a}=\frac{a}{\nu_g}, \quad \widetilde{\xi}=\frac{\xi}{\nu_g}.
\eeq
The constraint $P_{xx}^*+(d-1)P_{yy}^*=d$ leads to a cubic equation relating $\widetilde{\xi}$ and $\widetilde{a}$:
\begin{equation}
\label{3.18}
\widetilde{a}^2=d\widetilde{\xi}(1+2\tilde{\xi})^2.
\end{equation}
The real root of Eq.\ \eqref{3.18} gives the shear-rate dependence of $\widetilde{\xi}(\widetilde{a})$. It is given by
\beq
\label{3.19}
\widetilde{\xi}(\tilde{a})=\frac{2}{3}\sinh^2\left[\frac{1}{6}\cosh^{-1}\left(1+\frac{27}{d}\widetilde{a}^2\right)\right].
\eeq
For small shear rates, $\widetilde{\xi}\sim \widetilde{a}^2$ while $\widetilde{\xi}\sim \widetilde{a}^{2/3}$ for large shear rates \cite{GS03}.

Equations \eqref{3.14}--\eqref{3.16} along with Eq.\ \eqref{3.19} provide the shear-rate dependence of the elements of the (reduced) pressure tensor $P_{g,ij}^*$. These expressions (obtained from Grad's moment method) coincide with the exact results obtained for Maxwell molecules in the USF problem \cite{IT56}.  

In order to characterize the non-Newtonian transport properties, it is usual \cite{M89} to introduce the (dimensionless) nonlinear shear viscosity $\eta_g^*$ and the (dimensionless) first viscometric function $\Psi_g^*$ defined as
\beq
\label{3.20}
\eta_g^*(\widetilde{a})=-\frac{P_{g,xy}^*}{\widetilde{a}}, \quad \Psi_g^*(\widetilde{a})=\frac{P_{g,yy}^*-P_{g,xx}^*}{\widetilde{a}^2}.
\eeq
We recall again that the collisional moment $A_{g,ij}$ has been computed by neglecting nonlinear terms in the traceless pressure tensor $\Pi_{g,ij}$. This approximation yields $P_{g,yy}^*=P_{g,zz}^*$. The evaluation of $A_{g,ij}$ retaining all the quadratic terms in $\Pi_{g,ij}$ have been reported in Refs.\ \cite{TK95,ChVG15}. The inclusion of these nonlinear contributions to $A_{g,ij}$ allows us to obtain the normal stress differences 
$P_{g,yy}^*-P_{g,zz}^*$ in the plane orthogonal to the shear flow. However, given that computer simulations \cite{ChVG15} have shown that the above difference is in general very small, one can conclude that the expression \eqref{3.12} can be considered as accurate for obtaining the non-transport properties of the molecular gas.

Figure \ref{fig1} shows the shear-rate dependence of the rheological functions $\eta_g^*(\widetilde{a})$ and $\Psi_g^*(\widetilde{a})$ for a three-dimensional ($d=3$) molecular gas. Theoretical results are compared against computer simulation results obtained from the direct simulation Monte Carlo (DSMC) method  
\cite{B94}. As expected \cite{GS03}, $\eta_g^*$ and $|\Psi_g^*|$ are monotonically decreasing functions of the shear rate $\widetilde{a}$. A shear thinning effect is observed for the shear viscosity since the shearing inhibits the momentum transport: the actual value of $|P_{g,xy}^*|$ is smaller than the one predicted by Newton's law. With respect to the viscometric function, we see that it is negative and hence, $P_{g,xx}^*>P_{g,yy}^*$. This can interpreted as a breakdown of the energy equipartition. It is quite also apparent that the (approximate) theoretical results exhibit an excellent agreement with computer simulation results, even for very large values of the shear rate.

\section{Rheology of the granular gas}
\label{sec4}

Once the rheological properties of the molecular gas have been determined, the next step is to obtain the non-zero elements of the pressure tensor $P_{ij}$ defined by Eq.\ \eqref{2.21}. In the steady USF state, the (homogeneous) distribution function $f(\mathbf{V})$ of the granular gas obeys the kinetic equation
\beq
\label{4.1}
-\frac{\partial}{\partial V_i}\left(a_{ij}V_{j}+\xi V_i\right)f({\bf V})=J[f,f]+J_\text{BL}[f,f_g].
\end{equation}
Multiplying both sides of Eq.\ \eqref{4.1} by $m V_i V_j$ and integrating over velocity, one achieves the equation  
\begin{equation}
\label{4.2}
a_{ik}P_{kj}+a_{j k}{P}_{ki}+2\xi P_{ij}={B}_{ij}+C_{ij},
\end{equation}
where
\beq
\label{4.3} 
B_{ij}=m\int d{\bf V}\; V_i V_j  J[{\bf V}|f,f], 
\eeq
\beq
\label{4.4}
C_{ij}=m\int d{\bf V}\; V_i V_j J_\text{BL}[{\bf V}|f,f_g].
\eeq
As in the case of the molecular gas, the collisional moments $\mathsf{B}$ and $\mathsf{C}$ cannot be exactly evaluated. Therefore, we employ the same strategy as for $\mathsf{A}_{g}$; Grad's moment method \cite{G49} is considered to provide a good estimate of $\mathsf{B}$ and $\mathsf{C}$. In this approach, $f_g(\mathbf{V})$ is replaced by the distribution \eqref{3.9} while the granular gas distribution $f(\mathbf{V})$ is replaced by the trial distribution 
\beq
\label{4.5}
f({\bf V})\to f_{\text{M}}({\bf V})\left(1+\frac{m}{2 T} \Pi_{ij} V_i V_j\right),
\eeq
where  
\beq
\label{4.6}
\Pi_{ij}=\frac{P_{ij}}{p}-\delta_{ij}
\eeq
is the traceless part of the (dimensionless) pressure tensor $P_{ij}/p$ and 
\begin{equation}
\label{4.7}
f_{\text{M}}({\bf V})=n \left(\frac{m}{2\pi T}\right)^{d/2}\exp\left(-\frac{m V^2}{2 T}\right)
\end{equation}
is the Maxwellian distribution of the granular gas (i.e., defined at the temperature $T\neq T_g$).

The corresponding collision integrals defining the tensors $B_{ij}$ and $C_{ij}$ can be now computed by using the trial distributions \eqref{3.9} for $f_g$ and \eqref{4.5} for $f$. In this case and retaining only linear terms in $\boldsymbol{\Pi}_g$ and $\boldsymbol{\Pi}$, after some algebra the collisional moments $B_{ij}$ and $C_{ij}$ can be written as \cite{G02}
\beq
\label{4.8}
B_{ij}=-\nu  n T_g \left[\nu_\eta^* P_{ij}^*+\chi\left(\zeta^*-\nu_\eta^*\right) \delta_{ij}\right],
\eeq
\beqa
\label{4.9}
C_{ij}&=&-2 n T_g \frac{m}{m_g} \mu \left(\frac{1+\theta}{\theta}\right)^{3/2}\gamma \Big(X \delta_{ij}+ \chi^{-1} Y P_{ij}^*\nonumber\\
& & +Z P_{g,ij}^*\Big),
\eeqa
where $P_{ij}^*=P_{ij}/(n T_g)$, $\chi=T/T_g$ is the temperature ratio, and $\theta=m T_g/(m_g T)$ is the ratio of the mean square velocities of granular and molecular gas particles. Moreover, in Eqs.\ \eqref{4.8} and \eqref{4.9} we have introduced the quantities    
\begin{equation}
\label{4.10}
\nu=\frac{8\pi^{(d-1)/2}}{(d+2)\Gamma(d/2)}n\sigma^{d-1}\sqrt{\frac{T}{m}},
\end{equation}
\beq
\label{4.11}
\nu_\eta^*=\frac{3}{4d}(1+\al)\left(1-\al+\frac{2}{3}d\right), 
\eeq
\beq
\label{4.12}
\zeta^*=\frac{\zeta}{\nu}=\frac{d+2}{4d}(1-\al^2),
\eeq
\beq
\label{4.13}
\gamma=\frac{4\pi^{(d-1)/2}}{d\Gamma\left(\frac{d}{2}\right)}\left(\frac{m_g}{m}\right)^{1/2}\left(\frac{2T_g}{m}\right)^{1/2}n_g \overline{\sigma}^{d-1},
\eeq
\beq
\label{4.14}
X=
-\frac{\mu_g}{d+2}\left(d+\frac{\chi-1}{1+\theta^{-1}}\right),
\eeq
\beq
\label{4.15}
Y=
\frac{1}{(d+2)(1+\theta)}\Big[d \mu_g+\frac{(d+2)+(d+3)(\chi-1)\mu_g}{1+\theta^{-1}}\Big],
\eeq
\beq
\label{4.16}
Z=
\frac{\mu_g \theta}{(d+2)(1+\theta)}\Big[d+(d+3)\frac{\chi-1}{1+\theta^{-1}}\Big]-\frac{\theta}{(1+\theta)^2}.
\eeq
It is quite apparent that the expression of the drift or friction coefficient $\gamma$ is the same as the one obtained in the Brownian limit ($m/m_g\to \infty$) when the molecular gas is at equilibrium ($f_g=f_{g,\text{M}}$). In these conditions, a Kramers--Moyal expansion in the velocity jumps $\delta \mathbf{v}=(2/(1+m/m_g))(\widehat{{\boldsymbol {\sigma}}} \cdot {\mathbf g}_{12})\mathbf{g}_{12}$ allows to approximate 
the Boltzmann--Lorentz operator $J_\text{BL}[f,f_g]$ by a Fokker--Planck operator \cite{RL77,M89,BDS99,SVCP10}. Note that the expression \eqref{4.13} of $\gamma$ differs from the macroscopic Stokes law describing the Brownian motion of a massive particle in an equilibrium host fluid \cite{BHP94,BPH94}.


As usual in the USF problem for granular suspensions \cite{HTG17,HT19,THSG20}, we introduce the dimensionless shear rate
\beq
\label{4.19}
a^*=\frac{a}{\gamma},
\eeq
and Eq.\ \eqref{4.2} in dimensionless form can be rewritten as
\beqa
\label{4.20}
& & a_{ik}^*P_{kj}^*+a_{jk}^*{P}_{ki}^*+2\xi^* P_{ij}^*=-\nu^* \left[\nu_\eta^*P_{ij}^*+\chi (\zeta^*-\nu_\eta^*)\delta_{ij}\right]\nonumber\\
& & -2 \frac{m}{m_g} \mu \left(\frac{1+\theta}{\theta}\right)^{3/2}\left(X \delta_{ij}+ \chi^{-1} Y P_{ij}^*+Z P_{g,ij}^*\right),
\eeqa
where $\xi^*=\xi/\gamma$ and 
\beq
\label{4.21}
\nu^*=\frac{\nu}{\gamma}=\epsilon \chi^{1/2}, \quad
\epsilon=\frac{2^{d+2}d}{(d+2)\sqrt{\pi}}\phi \sqrt{T_g^*}.
\eeq
Here,
\beq
\label{4.22}
\phi=\frac{\pi^{d/2}}{2^{d-1}d \Gamma\left(\frac{d}{2}\right)}n\sigma^d
\eeq
is the solid volume fraction and
\beq
\label{4.23}
T_g^*=\frac{T_g}{m \sigma^2 \gamma^2}
\eeq
is the reduced background temperature. Note that the density dependence of $\nu^*$ arises from the scaling of the shear rate $a$ with the drift coefficient $\gamma$. If we had reduced the shear rate for instance with the collision frequency $\nu(T)$, then the above density dependence had been removed. On the other hand, since we want to study the non-Newtonian transport properties of the granular gas in thermal contact with a bath of elastic particles at temperature $T_g$ it seems natural to scale $a$ with respect to $\sqrt{T_g}$ instead of $\sqrt{T}$, as usual in dry granular gases \cite{SGD04}. Moreover, given that we want to make a close comparison with computer simulations \cite{HTG17,HT19,THSG20}, the theory must employ the same input parameters as in the simulation results. In this context, in order to express the rheological properties of the molecular gas in terms of the (reduced) shear rate $a^*$, one has to take into account the relations 
\beq
\label{4.24}
\widetilde{a}=\frac{\gamma}{\nu_g}a^*, \quad 
\widetilde{\xi}=\frac{\gamma}{\nu_g}\xi^*,
\eeq
where
\beq
\label{4.25}
\frac{\gamma}{\nu_g}=\frac{d+2}{\sqrt{2}d}\left(\frac{\overline{\sigma}}{\sigma_g}\right)^{d-1}\frac{m_g}{m}.
\eeq

The equation for the temperature ratio $\chi$ can be obtained from Eq.\ \eqref{4.20} as
\beq
\label{4.26}
2a^*P_{xy}^*+2 d \chi \xi^*+d \chi \left(\zeta^*+\zeta_g^*\right)=0,
\eeq
where $\zeta^*$ is defined by Eq.\ \eqref{4.12} and $\zeta_g^*=\zeta_g/\nu$ is given by   
\beq
\label{4.27}
\zeta_g^*=2 x (1-x^2)\mu^{1/2}\epsilon^{-1},
\eeq
with
\beq
\label{4.27.1}
x=\left(\mu_g+\mu \frac{T_g}{T}\right)^{1/2}.
\eeq
For elastic collisions ($\al=1$) and vanishing shear rates ($a^*=0$), $\zeta^*=\xi^*=0$, and Eq.\ \eqref{4.26} simply yields $\chi=1$ (energy equipartition), as expected. For inelastic collisions ($\al<1$) and arbitrary shear rates, Eq.\ \eqref{4.26} must be numerically solved.

The elements $P_{xx}^*$, $P_{yy}^*$, and $P_{xy}^*$ can be easily determined from Eq.\ \eqref{4.20}. In terms of the temperature ratio, they can be written as
\beq
\label{4.31}
P_{yy}^*=\frac{\Omega_{yy}}{\nu_{yy}}, \quad P_{xy}^*=\frac{\Omega_{xy}}{\nu_{yy}}-\frac{a^*}{\nu_{yy}}P_{yy}^*,
\eeq
\beq
\label{4.32}
P_{xx}^*=d \chi-(d-1)P_{yy}^*,
\eeq
where we have introduced the quantities
\beq
\label{4.33}
\nu_{yy}=\nu^* \nu_\eta^*+2\xi^*+2 \mu \theta \left(\frac{1+\theta}{\theta}\right)^{3/2} Y,
\eeq
\beq
\label{4.34}
\Omega_{yy}=-\nu^*\chi \left(\zeta^*-\nu_\eta^*\right)-2\frac{m}{m_g}\mu \left(\frac{1+\theta}{\theta}\right)^{3/2}\left(X+Z P_{g,yy}^*\right),
\eeq
\beq
\label{4.35}
\Omega_{xy}=-2\frac{m}{m_g}\mu \left(\frac{1+\theta}{\theta}\right)^{3/2} Z P_{g,xy}^*.
\eeq

The temperature ratio can be determined by substituting Eqs.\ \eqref{4.27} and \eqref{4.31} into Eq.\ \eqref{4.26}. This yields the equation
\beq
\label{4.35a}
a^{*}=\sqrt{\frac{d}{2}\frac{2\mu^2 \left(\frac{1+\theta}{\theta}\right)^{1/2}(1-\chi)-\left(\nu^* \zeta^*+2\xi^*\right)\chi}{\frac{\Omega_{xy}/a^*}{\nu_{yy}}-\frac{\Omega_{yy}}{\nu_{yy}^2}}}.
\eeq
As occurs in previous works on granular suspensions \cite{HTG17,HT19,HT19,THSG20}, the temperature ratio $\chi$ cannot be expressed in Eq.\ \eqref{4.35a} as an explicit function of the (reduced) shear rate $a^*$ and the remaining parameters of the system. However, the dependence of $\chi$ on the latter parameters can implicitly be obtained from the physical solution to Eq.\ \eqref{4.35a} as $a^{*2}(\chi,\al,\sigma/\sigma_g,m/m_g,\phi, T_g^*)$.

As for the molecular gas, non-Newtonian transport properties of the granular gas  are characterized by the nonlinear shear viscosity
\beq
\label{4.38}
\eta^*=-\frac{P_{xy}^*}{a^*},
\eeq
and the normal stress difference 
\beq
\label{4.39}
\Psi^*=P_{xx}^*-P_{yy}^*.
\eeq
As expected, both quantities are highly \emph{nonlinear} functions of the (reduced) shear rate $a^*$ and the parameter space of the system.  


\subsection{Navier--Stokes shear viscosity}

Before considering the shear-rate dependence of the non-Newtonian transport properties of the granular suspension, it is quite instructive to consider the limit of small shear rates (Navier--Stokes regime) when the molecular gas is at equilibrium. This was the situation analyzed in Ref.\ \cite{GG22a}.

For vanishing shear rates ($a^*=0$), $\xi^*=0$ and Eq.\ \eqref{4.20} becomes $\zeta^*+\zeta_g^*=0$. This yields the following cubic equation for $x$:
\beq
\label{4.28}
2x(x^2-1)=\vartheta, \quad \vartheta=\frac{d+2}{4d}\mu^{-1/2}\epsilon(1-\al^2).
\eeq
The physical root of Eq.\ \eqref{4.28} can be written as \cite{S03a}
\begin{widetext}
\beq
\label{4.29}
x=\Bigg\{
\begin{array}{cc}
\frac{\sqrt{3}}{3}\left\{\sqrt{3}\cos\left[\frac{1}{3}\sin^{-1}\left(\frac{3\sqrt{3}}{4}\vartheta\right)\right]+
\sin\left[\frac{1}{3}\sin^{-1}\left(\frac{3\sqrt{3}}{4}\vartheta\right)\right]\right\},&\vartheta \leq \frac{4\sqrt{3}}{9}\\
\frac{2\sqrt{3}}{3}\cosh\left[\frac{1}{3}\cosh^{-1}\left(\frac{3\sqrt{3}}{4}\vartheta\right)\right],&\vartheta \geq \frac{4\sqrt{3}}{9}.
\end{array}
\eeq
\end{widetext}
Thus, according to Eq.\ \eqref{4.27.1}, the temperature ratio $T/T_g$ is 
\beq
\label{4.30}
\frac{T}{T_g}=\frac{m/m_g}{\left(1+\frac{m}{m_g}\right)x^2-1}.
\eeq
Equations \eqref{4.29} and \eqref{4.30} agree with those obtained in Ref.\ \cite{GG22a}.

In the Navier--Stokes domain, $P_{g,ij}^*=\delta_{ij}$ and since $\xi^*\propto \widetilde{\xi}\sim a^{*2}\to 0$ when $a^*\to 0$, then Eqs.\ \eqref{4.31}--\eqref{4.35} lead to $P_{xx}^\text{NS}=P_{yy}^\text{NS}=n T$, as expected. In addition, the shear stress 
\beq
\label{4.30.1}
P_{xy}^\text{NS}=-\eta_\text{NS} a,
\eeq
where the Navier--Stokes shear viscosity $\eta_\text{NS}$ can be written as 
\beq
\label{4.36}
\eta_\text{NS}=\frac{\eta_0}{\nu_\eta^*+\widetilde{\nu}_\eta},
\eeq
where 
\beq
\label{4.36.1}
\eta_0=\frac{(d+2)\Gamma(d/2)}{8\pi^{(d-1)/2}}\sigma^{1-d}\sqrt{mT} 
\eeq
is the Navier--Stokes shear viscosity of an ordinary dilute gas of hard spheres ($\al=1$), $\nu_\eta^*$ is given by Eq.\ \eqref{4.11} and 
\beq
\label{4.37}
\widetilde{\nu}_\eta=2\frac{m}{m_g}\mu \left(\frac{1+\theta}{\theta}\right)^{3/2}Y \epsilon^{-1}\chi^{-3/2}.
\eeq
The expression \eqref{4.36} of $\eta_\text{NS}$ agrees with the one derived in Ref.\ \cite{GG22a} in the low-density regime from the Chapman--Enskog method. This shows the self consistency
of the present results with those reported in the Navier--Stokes limiting case.


\section{Brownian limiting case}
\label{sec5}

It is quite apparent that the results obtained in Sec.\ \ref{sec4} apply for arbitrary values of the diameter and mass ratios. However, in the case of particle-laden suspensions \cite{S20} where collisions play an essential role in the dynamics of grains, the granular particles are much heavier than the particles of the surrounding molecular gas ($m/m_g\to \infty$). This is referred here as to the Brownian limit. 

As discussed in the Introduction section, given the technical difficulties for describing a system constituted by two or more phases (gas-solid flows), a coarse-grained approach is 
usually adopted for analyzing gas-solid flows. In this approach, the influence of gas-phase on grains is incorporated in the starting kinetic equation via an effective fluid-solid interaction force \cite{K90,G94,J00}. Based on the results obtained in direct numerical simulations, for high-velocity (but low-Reynolds numbers) gas-solid flows, the effective force is constituted by two terms in the USF problem: (i) a drag force term proportional to the particle velocity and (ii) a stochastic Langevin-like term. While the first term models the energy lost by grains due to its friction on the viscous gas, the second term accounts for the energy transferred to the solid particles due to their collisions with the particles of the interstitial molecular gas. Thus, in the low-density regime, the starting kinetic equation for studying the USF is the Boltzmann equation plus a Fokker--Planck term. In fact, this latter term can be formally obtained from the Boltzmann--Lorentz collision operator in the limiting Brownian case when the molecular gas is at equilibrium. However, some papers \cite{RSD83,PZ23} have shown that the conventional Langevin equation is modified by the shear flow so that, these new contributions should be incorporated in the coarse-grained description when one analyzes the USF problem in granular suspensions.   

 \begin{figure}[ht]
\centering
\includegraphics[width=0.4\textwidth]{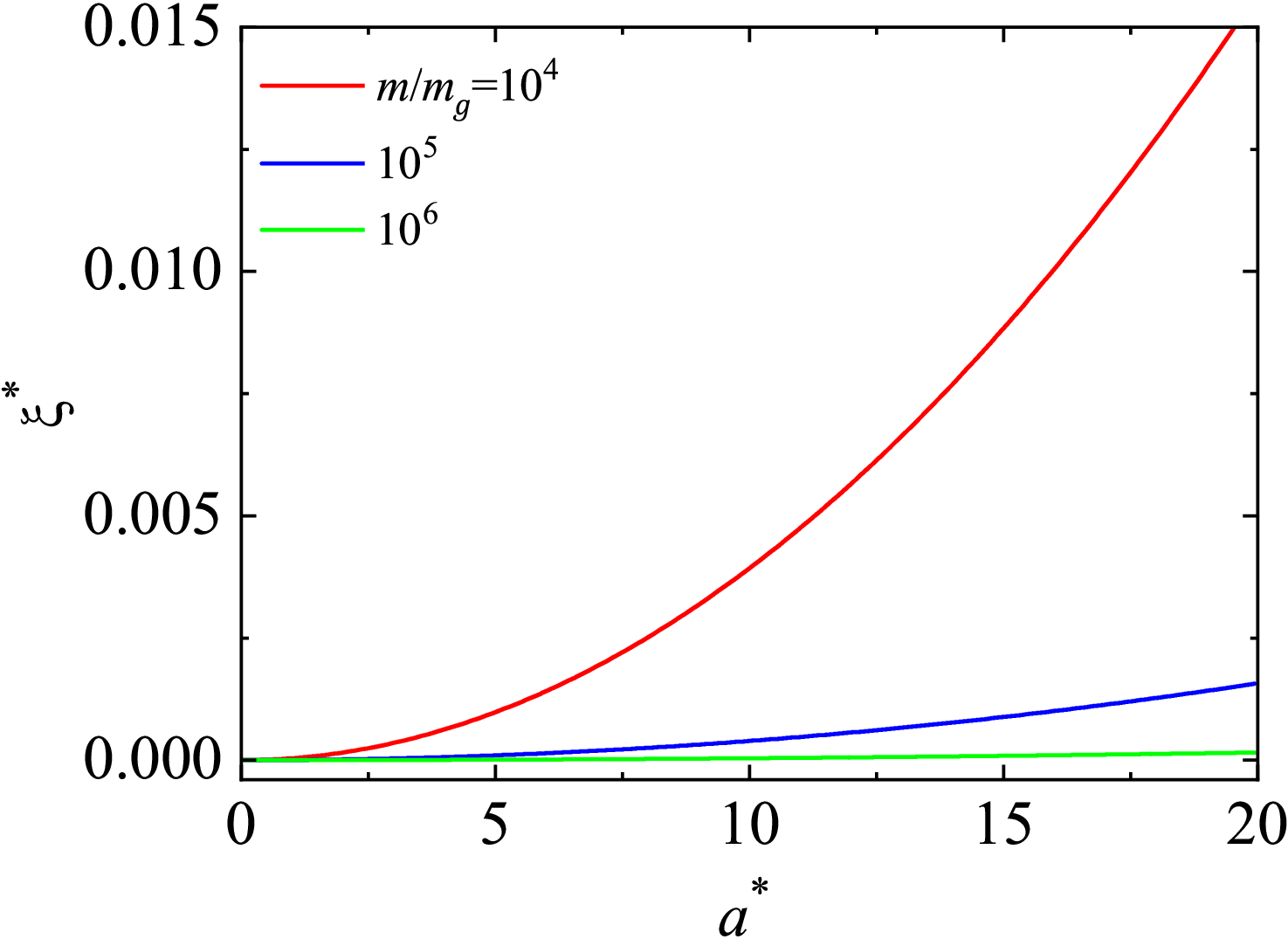}
\caption{Thermostat parameter $\xi^*=\xi/\gamma$ versus the (reduced) shear rate $a^*=a/\gamma$ for hard spheres ($d=3$) with $\sigma=\sigma_g$ and three different values of the mass ratio $m/m_g$.}
\label{fig2}
\end{figure}

The objective of this section is to assess the consistency between the results obtained here (which are based on a collisional model) in the Brownian limit with those previously obtained \cite{HTG17,HT19,THSG20} by employing the Langevin equation. An important conclusion of those papers is 
that the non-Newtonian viscosity $\eta^*$ exhibits an $S$ shape in a plane of stress–strain rate (discontinuous shear thickening effect).    

At a given \emph{finite} value of the (reduced) shear rate $a^*$, according to Eqs.\ \eqref{4.24} and \eqref{4.25}, $\widetilde{a}\propto m_g/m$ and $\widetilde{\xi}\propto m_g/m$. This means that in the Brownian limiting case ($m/m_g\to \infty$) the shear rate $a$ in units of the collision frequency $\nu_g$ of the molecular gas is very small. Consequently, the pressure tensor of the granular gas behaves as $P_{g,ij}^*=\delta_{ij}$ and hence, the granular gas does not perceive the presence of a strong shear field. Although not explicitly stated, this is in fact one of the assumptions in the suspensions models employed by starting from the conventional Langevin equation in the USF state \cite{HTG17,HT19,THSG20,KIB14,TK95,SMTK96,H13,SA17,SA20}.

In the Brownian limit, $\mu_g\to 0$, $\mu \to 1$, $\chi\equiv \text{finite}$, and so $\theta \sim m/m_g \chi^{-1}$, $X+Z\sim -m_g/m$, and $Y\sim \theta^{-1}\sim (m_g/m)\chi$. In addition, the thermostat parameter $\xi^*=(\nu_g/\gamma)\widetilde{\xi}$ is negligibly small when $m\gg m_g$. This is illustrated in Fig.\  \ref{fig2} where $\xi^*$ is plotted versus $a^*$ for $d=3$, $\sigma=\sigma_g$, and different values of the mass ratio. Since we see in Fig.\  \ref{fig2} that $\xi^*\to 0$ as $m/m_g\to \infty$, we can take $\xi^*=0$ in Eqs.\ \eqref{4.31}--\eqref{4.35a}. Thus, after taking carefully the Brownian limit ($m/m_g\to \infty$) in the above equations, one achieves the expressions
\beq
\label{5.1}
P_{yy}^*=\frac{2-\nu^* \chi \left(\zeta^*-\nu_\eta^*\right)}{2+\nu^* \nu_\eta^*},
\eeq
\beq
\label{5.2}
P_{xy}^*=-\frac{a^*}{2+\nu^* \nu_\eta^*}P_{yy}^*,
\eeq
\beq
\label{5.3}
P_{xx}^*=\frac{2\left[d(\chi-1)+1\right]+\nu^*\chi\left[(d-1)\zeta^*+\nu_\eta^*\right]}{2+\nu^* \nu_\eta^*},
\eeq
\beq
\label{5.4}
a^*=\sqrt{\frac{d}{2}\frac{\nu^* \zeta^*+2\left(1-\chi^{-1}\right)}{\nu^* \left(\nu_\eta^*-\zeta^*\right)+2\chi^{-1}}}\left(2+\nu^*\nu_\eta^*\right).
\eeq
Equations \eqref{5.1}--\eqref{5.4} are consistent with the results obtained in Refs.\ \cite{HTG17,HT19,GGG19} from a coarse-grain approach.

In summary, we can conclude that the expressions of the rheological properties derived here from a suspension model that accounts for the grain-gas collisions reduce in the Brownian limit ($m\gg m_g$) to those obtained in previous works \cite{HTG17,HT19,GGG19}  when the influence of the interstitial gas on grains is accounted for by means of an effective  external force. This achievement justifies the use of the conventional Langevin-like model for determining the rheology of granular suspensions for arbitrary shear rates.

\section{Some illustrative systems. Comparison between theory and computer simulations}
\label{sec6}

\begin{figure}[ht!]
\centering
\includegraphics[width=0.4\textwidth]{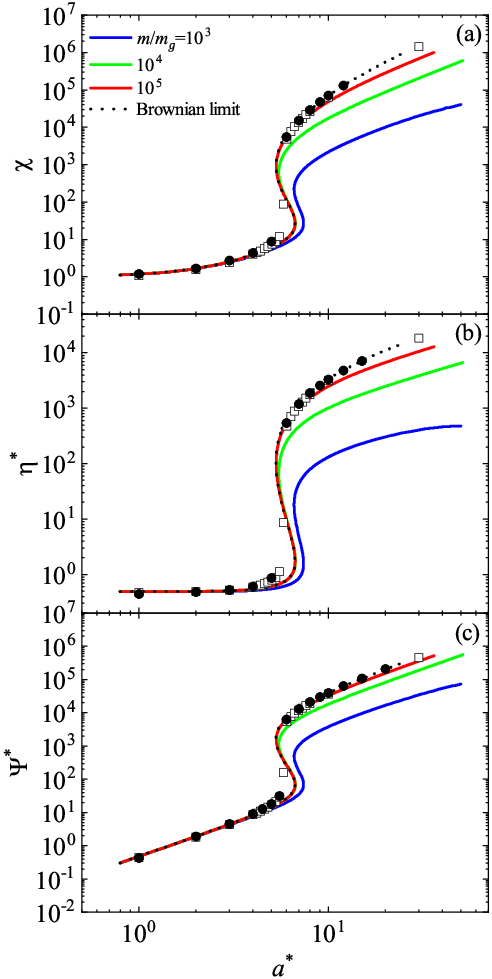}
\caption{Plots of the (reduced) granular temperature $\chi$
[panel (a)], the non-Newtonian shear
viscosity $\eta^*$ [panel (b)], and the viscometric function $\Psi^*$ [panel (c)] as a function
of the (reduced) shear rate $a^*$ for $\al=0.9$ and different values of the mass ratio $m/m_g$: $10^3$ (blue lines), $10^4$ (green lines), $10^5$  (red lines), and the Brownian limiting case (dotted lines). Here, $T^*_g=1$ , $d=3$, and $\phi=0.0052$. The filled symbols refer to DSMC simulations carried out in this paper, while the open ones refer to event-driven Langevin simulations performed in Ref.\ \cite{HTG17}. Both simulations have been made in the Brownian limit.}
\label{fig3}
\end{figure}

\begin{figure}[ht!]
\centering
\includegraphics[width=0.4\textwidth]{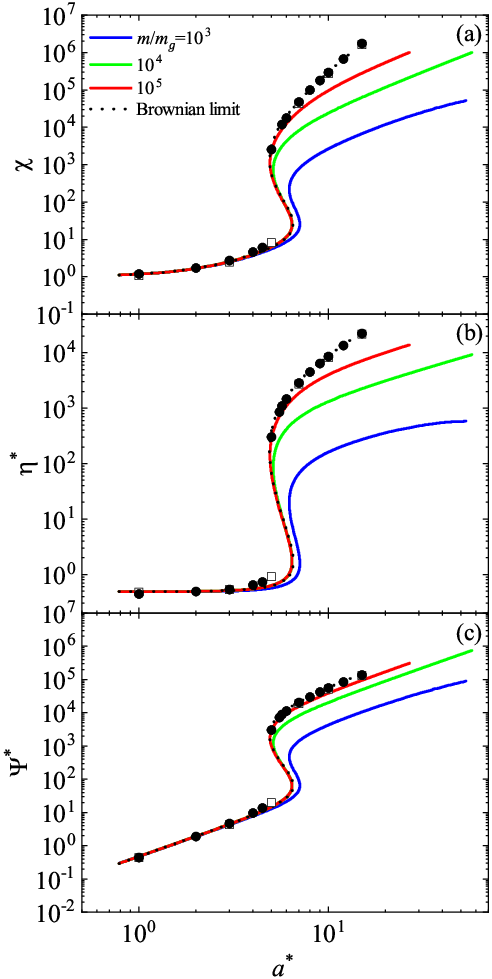}
\caption{Plots of the (reduced) granular temperature $\chi$
[panel (a)], the non-Newtonian shear
viscosity $\eta^*$ [panel (b)], and the viscometric function $\Psi^*$ [panel (c)] as a function
of the (reduced) shear rate $a^*$ for $\al=1$ and different values of the mass ratio $m/m_g$: $10^3$ (blue lines), $10^4$ (green lines), $10^5$  (red lines), and the Brownian limiting case (dotted lines). Here, $T^*_g=1$ , $d=3$, and $\phi=0.0052$. The filled symbols refer to DSMC simulations carried out in this paper, while the open ones refer to event-driven Langevin simulations performed in Ref.\ \cite{HTG17}. Both simulations have been made in the Brownian limit.} 
\label{fig4}
\end{figure}

The dependence of the (reduced) temperature $\chi$, the non-Newtonian shear viscosity $\eta^*$, and the viscometric function $\Psi^*$ on the (reduced) shear rate $a^*$ is shown in Figs. \ref{fig3} and \ref{fig4} for inelastic ($\alpha=0.9$) and elastic ($\alpha=1$) collisions, respectively. Apart from studying the shear-rate dependence of the rheological properties, our main goal here is to assess the impact of the mass ratio $m/m_g$ on them. Additionally, Eq. \eqref{4.25} introduces the diameter ratio $\sigma/\sigma_g$ as a ``new'' parameter for consideration. To ensure that the effect of inelastic collisions is on par with that of the (elastic) Boltzmann-Lorentz collision operator, the (reduced) background temperature $T_g^*\propto \nu/\gamma$ is chosen to be constant and approximately close to unity. In this specific instance, $T_g^*=1$. The selection of $T_g^*$ as a free parameter imposes the following constraint between the diameter ratio ($\sigma/\sigma_g$) and mass ratio ($m/m_g$):
\beq
\label{6.1}
\frac{\sigma_g}{\sigma}=\left[\left(\frac{\sqrt{\pi}}{4\sqrt{2}}\frac{n}{n_g}\sqrt{\frac{m}{m_g}}\frac{1}{\phi\sqrt{T_g^*}}\right)^{1/(d-1)}-1\right].
\eeq
As mentioned in Sec.\ \ref{sec4}, an explicit density dependence emerges due to the scaling of the background temperature with the known quantity $\gamma$. In this paper, we consider a very dilute gas with $\phi=0.0052$. Furthermore, since the granular gas is present in tracer concentration, the number density of granular particles $n_g$ is chosen to be an order of magnitude greater than the number density of the molecular gas $n$. More specifically, we take $n/n_g=10^{-3}$. It is important to highlight that Eq. \eqref{6.1} ensures $\xi^*\to 0$ as $m/m_g\to \infty$, thereby maintaining the consistency of Eqs. \eqref{4.31}--\eqref{4.35a} with those expressions derived from the Fokker--Planck operator \cite{GG22a} in the Brownian limit ($m/m_g\to \infty$).

The principal conclusion drawn from Figs.\ \ref{fig3} and \ref{fig4} is that variations in the mass ratio $m/m_g$ do not significantly alter the observed trends in prior studies \cite{HTG17,GGG19,GG20} within a coarse-grained approach. Notably, there is a discontinuous transition in all rheological properties with an increase in shear rate. Specifically, Panel (b) of Figs.\ \ref{fig3} and \ref{fig4} underscores the presence of DST in shear viscosity $\eta^*$ under sufficiently large mass ratios. Namely, the non-Newtonian shear viscosity $\eta^*$ discontinuously increases/decreases (at a certain value of $a^*$) as the (scaled) shear rate gradually increases/decreases. Quantitatively, it becomes apparent that, for a constant value of $a^*$, larger mass ratios marginally increase the magnitude of the $\eta^*$ discontinuity. The application of a shear flow impels particles to move together at the same velocity $U_x$. Over time, collisions between particles in each layer of the flow induce momentum transmission to the $y$ axis. Heavier particles are provided with more kinetic energy to maintain a constant velocity profile, leading to increased viscosity and temperature, as evident in Panel (a) of Figs.\ \ref{fig3} and \ref{fig4}. 

Analytically, Eq.\ \eqref{4.25} indicates that $\widetilde{a}\to 0$ when $m/m_g\to \infty$ for a finite value of $a^*$. Thus, Fig.\ \ref{fig1}(a) asserts that the viscosity of the molecular gas increases. Additionally, the balance between collisional cooling due to the inelasticity of grain collisions and bath thermalization depends solely on temperature, as $T_g^*$ is held constant. However, an increase in the mass ratio $m/m_g$ elevates temperature due to the higher kinetic energy required to move the entire system at the same velocity. Consequently, the collision frequency $\nu$ rises, leading to more collisions and facilitating momentum transmission. This effect combined with the increased viscosity of the molecular gas causes the viscosity of the suspension to increase. Although similar trends are noted for $\Psi^*$, it is noteworthy that the combined influence of $m/m_g$ and $a^*$ on the viscometric function $\Psi^*$ is substantial. While this quantity remains negligible for low shear rates, it abruptly increases for moderately small shear rates (e.g., $a^*\approx 0.1$). 

Theoretical predictions are compared against molecular dynamics (event-driven) simulations performed in Ref.\ \cite{HTG17} and DSMC results carried out in this paper. Both kind of simulations have been restricted to the Brownian limiting case 
($m/m_g\to \infty$). Some technical details of the application of the DSMC method to the system studied in this paper are provided in the supplementary material of Ref.\ \cite{GG22a} available at \url{https://doi.org/10.1017/jfm.
2022.410}. The absence of simulation data with finite mass ratios arises from the condition that, despite the granular gas being rarefied, collisions between grains are considered. Thus, given that the molecular and granular gases evolve on significantly different time scales, this difference becomes more accentuated when the mass ratio is very large. This means that it is quite difficult in the simulations to have a number of grain-grain collisions comparable to the ones occur between molecular and granular particles when $m/m_g \gg 1$. On the other hand, for smaller mass ratios, evolution
time scales for molecular and granular gases become comparable, enabling the validation of theoretical results with DSMC data [see Fig. 8].

It is quite apparent that the agreement between theory and computer simulations is generally favorable, with the exception of a localized region proximate to the transition point. In this region, molecular dynamics (MD)  simulation data indicate a more pronounced transition compared to the Boltzmann results. Given that the selected reference frequency is $\gamma$, it is conjectured that this slight disparity is primarily attributable to the marginal density dependence of the (reduced) collision frequency $\nu^*=\nu/\gamma$. This effect is not effectively accounted for in the Boltzmann kinetic theory.

\begin{figure}[ht]
\centering
\includegraphics[width=0.4\textwidth]{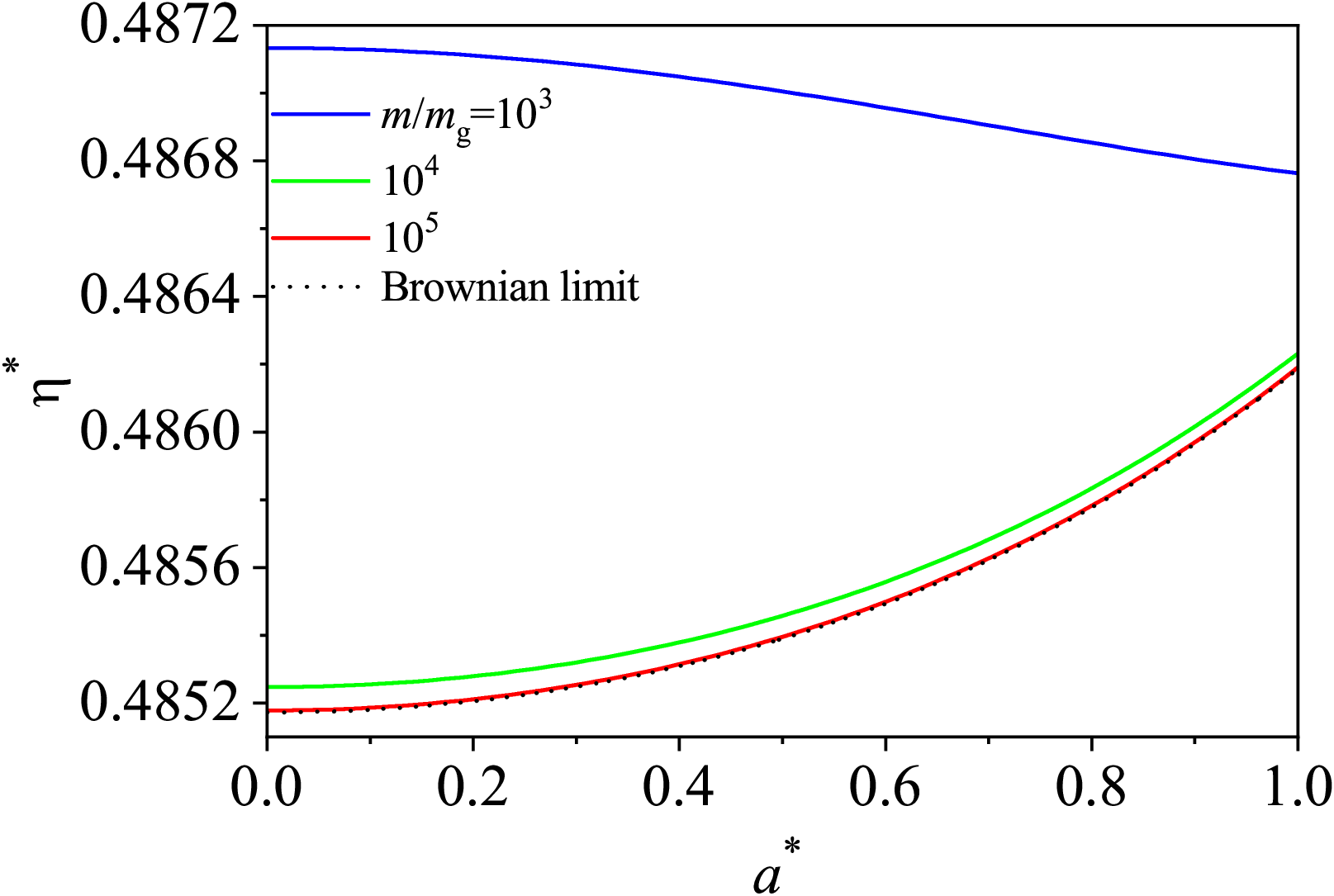}
\caption{Plot of the non-Newtonian shear
viscosity $\eta^*$ as a function
of the (reduced) shear rate $a^*$ for $\al=0.9$ and different values of the mass ratio $m/m_g$: $10^3$ (blue line), $10^4$ (green line), $10^5$  (red line), and the Brownian limiting case (dotted line). Here, $T^*_g=1$ , $d=3$, and $\phi=0.0052$.}
\label{fig5}
\end{figure}
\begin{figure}[ht]
\centering
\includegraphics[width=0.4\textwidth]{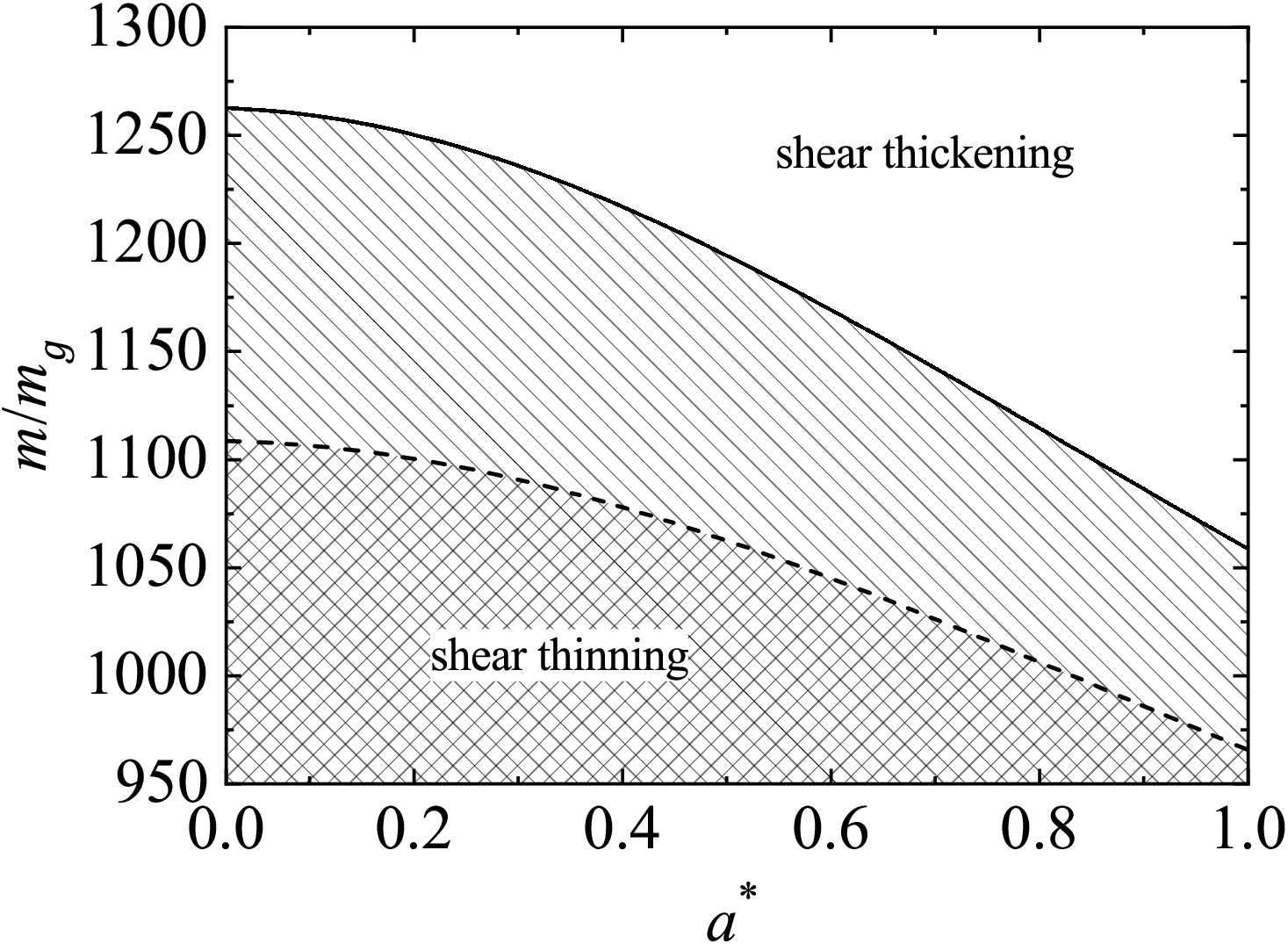}
\caption{Phase diagram of the shear thinning to thickening transition as a function
of the mass ratio $m/m_g$ and the (reduced) shear rate $a^*$ for $\alpha=0.9$ (solid line) and $\alpha=1$ (dashed line). The hatched regions correspond to the combined values of $a^*$ and $m/m_g$ where shear thinning effect occurs. 
Here, $T^*_g=1$ , $d=3$, and $\phi=0.0052$.}
\label{fig5a}
\end{figure}
\begin{figure}[ht]
\centering
\includegraphics[width=0.4\textwidth]{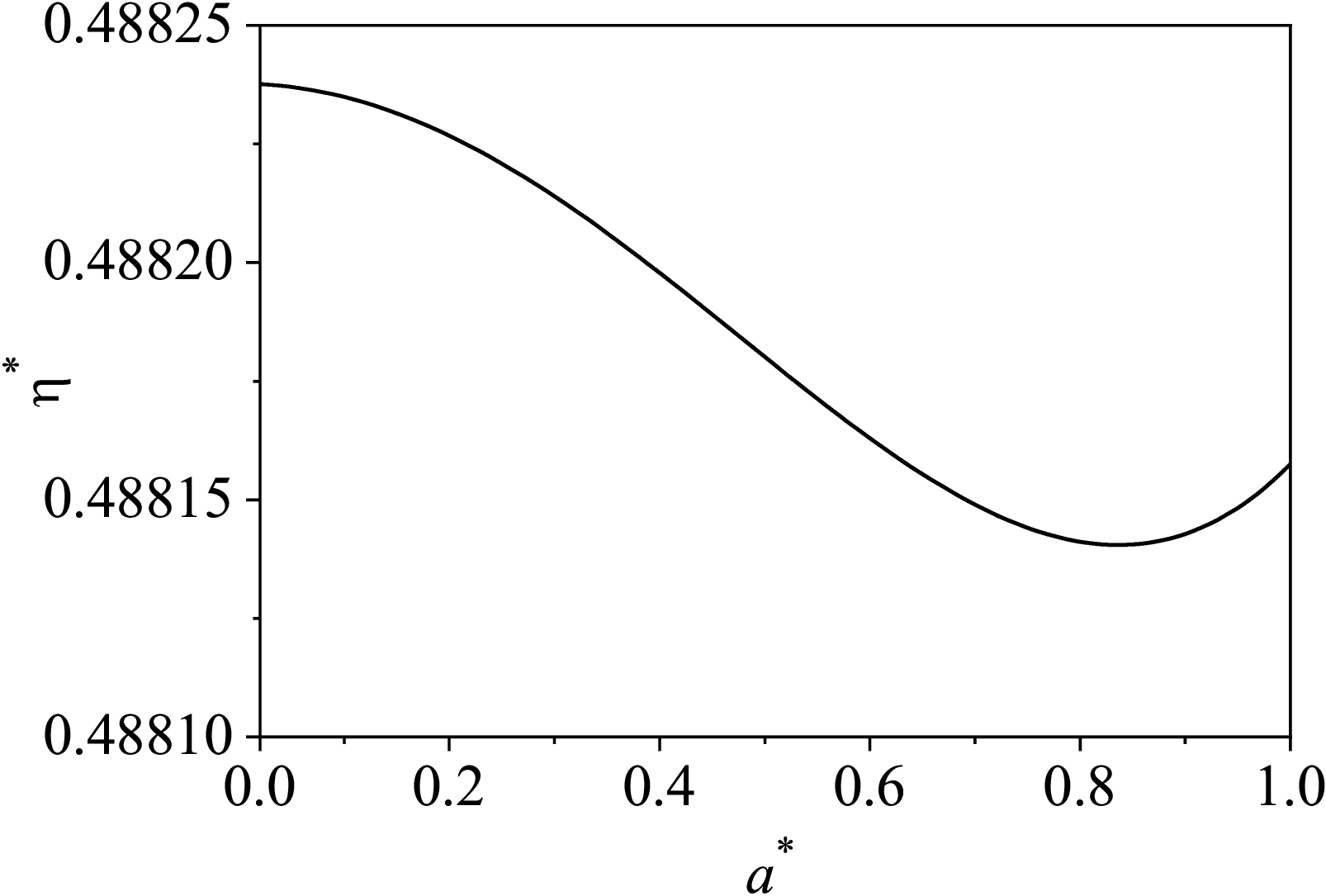}
\caption{Plot of the non-Newtonian shear
viscosity $\eta^*$ as a function
of the (reduced) shear rate $a^*$ for $\al=1$ and $m/m_g=10^3$. Here, $T^*_g=1$, $d=3$, and $\phi=0.0052$.}
\label{fig5b}
\end{figure}

In the absence of shear or with minimal shear, no such amount of energy is required to move the fluid at a constant velocity $U_x$. Pure collisional effects from the isolated action of the background fluid and inelasticity emerge. As granular particles become heavier, the inertial effect of mass causes the grains to necessitates more collisions with the background fluid for thermalization. This produces a decrease in granular temperature and viscosity, as observed in Fig.\ \ref{fig5} for small values of the (reduced) shear rate $a^*$. Remarkably, shear-thinning effects are observed for sufficiently small values of both $m/m_g$ and $a^*$. To complement Fig.\ \ref{fig5}, Fig.\ \ref{fig5a} illustrates the phase diagram depicting the critical mass ratio at which the sign of the slope $\partial \eta^*/\partial a^*$ changes, indicating a transition from shear thinning to shear thickening. It is evident from the diagram that, at a given value of the (reduced) shear rate $a^*$, an increase in the mass ratio $m/m_g$ leads to a more pronounced rise in the non-Newtonian viscosity $\eta^*$. By examining the variation of $a^*$, it becomes clear that, even for the same value of $m/m_g$, the suspension may exhibit either shear thinning or shear thickening behavior depending on the specific value of $a^*$, as demonstrated in Fig.\ \ref{fig5b} for $m/m_g=10^3$. A similar transition was observed in moderately dense molecular gases (see, for instance, Fig.\ 1 of Ref.\ \cite{SMDB98}). In this case, for moderate densities, 
the collisional contribution to the nonlinear shear viscosity is responsible for a crossover from shear thinning to 
shear thickening at sufficiently large shear rates. In the granular suspension studied here, the increase in density results in an increase of the number of collisions.  Therefore, according to Eq.\ (59) and for fixed values of $\phi$ and $n/n_g$, as the mass ratio $m/m_g$ increases the diameter ratio $\sigma/\sigma_g$ also increases. Given that $n$ is constant, an increase in $\sigma$ resembles an increase in density since $\phi\propto n\sigma^d$, thereby increasing the number of collisions suffered by the granular particles. This increase (caused by both the density and the diameter of the spheres) causes the transmission of momentum between the different layers of the fluid to be more effective and thus causes the viscosity to increase with increasing the mass ratio.

\begin{figure}[ht]
\centering
\includegraphics[width=0.4\textwidth]{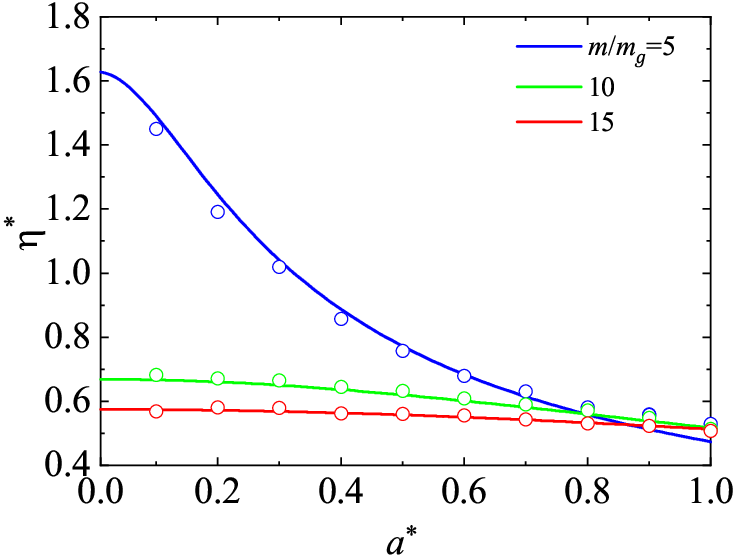}
\caption{Plot of the non-Newtonian shear
viscosity $\eta^*$ as a function
of the (reduced) shear rate $a^*$ for $\al=0.9$ and different values of the mass ratio $m/m_g$: $5$ (red line and symbols), $10$ (green line and symbol), and $15$  (red lines and symbols). Here, $T^*_g=1$ , $d=3$, and $\phi=0.0052$. The symbols refer to DSMC simulations.}
\label{fig5c}
\end{figure}

To assess the accuracy of Grad's expression for the non-Newtonian shear viscosity in the region $a^*\leq 1$ (where non-Newtonian effects are still important) and relatively small mass ratios, Fig.\ \ref{fig5c} shows the shear-rate dependence of $\eta^*$
for $d=3$, $T_g^*=1$, $\phi=0.0052$, and three different values of $m/m_g$. Theoretical results are compared against DSMC simulations. With this purpose and to avoid an enormous difference between the evolution times of each phase (molecular and granular gases), we consider $n/n_g=10^{-2}$ in Eq.\ \eqref{6.1}. As expected from the phase diagram \ref{fig5a}, a shear thinning effect is always found for all the mass ratios considered.
For not large shear rates, $\eta^*$ decreases with increasing the mass ratio $m/m_g$ at a given value of $a^*$. However, this tendency changes as the shear rate increases since there is a crossing between the different curves for sufficiently large values of $a^*$. Figure \ref{fig5c} highlights again the good agreement found between theory and computer simulations.

\begin{figure}[ht!]
\centering
\includegraphics[width=0.4\textwidth]{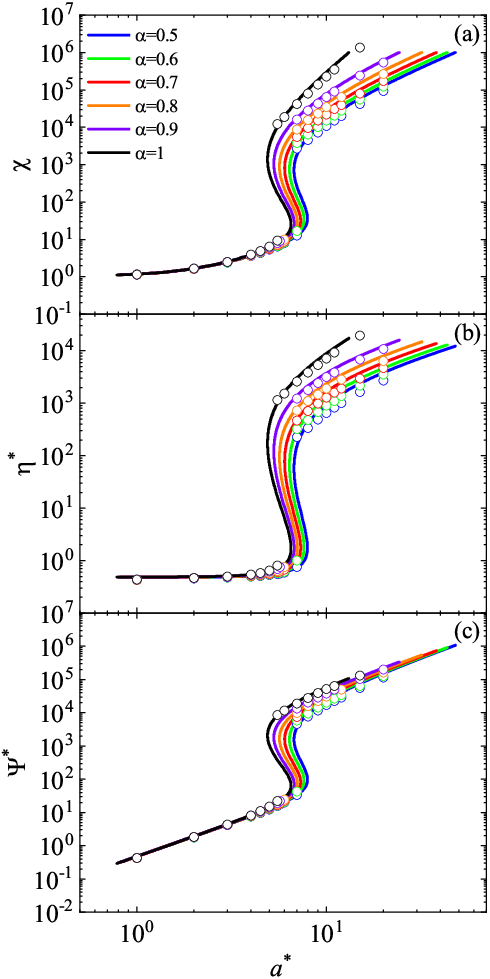}
\caption{Plots of the (reduced) granular temperature $\chi$
[panel (a)], the non-Newtonian shear
viscosity $\eta^*$ [panel (b)], and the viscometric function $\Psi^*$ [panel (c)] in the Brownian limit as a function
of the (reduced) shear rate $a^*$ for different values of $\alpha$: $0.5$ (blue lines), $0.6$ (green lines), $0.7$  (red lines), $0.8$  (orange lines), $0.9$  (purple lines), and $1$  (black lines). Here, $T^*_g=1$ , $d=3$, and $\phi=0.0052$. The symbols refer to DSMC simulations.}
\label{fig6}
\end{figure}

Another important effect to consider is the impact of inelasticity on rheology. To explore this, our focus is on the Brownian limit where the influence of $\alpha$ is isolated from the mass ratio. Figure \ref{fig6} illustrates the rheological properties for various values of $\alpha$. Firstly, it is observed that inelasticity attenuates the DST transition in the non-Newtonian viscosity $\eta^*$. This is primarily attributed to the collisional cooling resulting from interactions between grains, leading to a decrease in the (reduced) temperature $\chi$. As the most noticeable effect, this temperature decrease reduces collision frequency. Consequently, momentum transfer becomes more challenging, causing a viscosity decrease for a given $a^*$. We also see that the effect of $\alpha$ on the viscometric function $\Psi^*$ is smaller than the one found for the temperature $\chi$ and the non-Newtonian viscosity $\eta^*$. Furthermore, a satisfactory level of agreement between theory and DSMC simulations is demonstrated, although some discrepancies between both approaches are observed. These differences probably arise from the influence of the fourth-degree velocity moments on the distribution function of the granular gas as the shear rate increases. Note that Grad's solution \eqref{3.9} neglects the above moments since it includes only the second-degree velocity moment (shear stress).

Apart from comparing theory with computer simulations, it would be desirable to assess the usefulness of the theoretical predictions offered by the suspension model considered in this paper via a comparison with experimental results. However, given that our results are restricted to very dilute systems, it is not quite usual to find in the granular literature experiments carried for such range of densities. In fact, we are not aware of any work where the rheological properties of \emph{dilute} gas-solid flows have been measured. In any case, the question then is to what extent systems subject to the restrictions considered here  can be found in nature and replicated in the laboratory. In a recent paper where two of the authors of the present work have been involved \cite{GASG23}, a system (a suspension of gold grains immersed in hydrogen molecular gas at normal temperature and pressure, see Table I of Ref.\ \cite{GASG23}) has been proposed where grain-grain collisions can play a relevant role in the evaluation of the transport coefficients of the suspension. We expect that the results reported in this paper stimulate the performance of these sort of experiments.

\section{Summary and concluding remarks}
\label{sec7}
In recent years, there has been a growing interest in employing kinetic theory tools to determine the non-Newtonian transport properties of inertial suspensions under simple shear flow. Numerous studies, drawing on the Boltzmann kinetic equation (applicable to very dilute systems) and/or the Enskog kinetic equation (suitable for moderately dense systems), have made substantial contributions to this field. In particular, several works \cite{TK95, SMTK96, GTSH12, ChVG15, HTG17, SA20, HT19, GGG19, GG20} have derived explicit expressions for the shear-rate dependence of the kinetic temperature, non-Newtonian viscosity, and viscometric functions.

Among various gas-solid flows, the particle-laden suspensions \cite{G94,J00,KH01,TS14,FH17,W20,LGW20} pose an intriguing problem. In such suspensions, dilute and immiscible particles are immersed in a denser fluid, causing hydrodynamic interactions to become less relevant \cite{S20,B72}. Consequently, the dynamics of solid particles are influenced by thermal fluctuations in the fluid, with interparticle, external, and Brownian forces prevailing \cite{BB88}. Assuming the number density of the solid phase is significantly smaller than that of the surrounding fluid, the latter remains practically unaffected by the presence of the solid particles. In this scenario, the external fluid may be treated as a thermostat (or bath) at a fixed temperature $T_g$. Thus, gas-solid interactions are effectively modeled via a fluid-solid force in the starting kinetic equation \cite{K90,G94,J00}.

Some models for granular suspensions \cite{LMJ91,TK95,SMTK96,WZL09,PS12,H13,WGZS14,SA17,ASG19,SA20} consider only isolated body resistance through a linear drag law, while others \cite{PLMPV98,GTSH12} include an additional Langevin-type stochastic term. Analytical results from these models, validated against computer simulations, generally exhibit good agreement under conditions of practical interest. However, questions arise about whether these models accurately reproduce real situations and if they can be recovered using purely kinetic arguments.

In this work, we have started from a suspension model that consider real collisions between grains and particles of the background gas in the starting kinetic equation. The influence of the gas phase on solid particles is introduced via the (elastic) Boltzmann-Lorentz collision operator $J_g[f,f_g]$. As in previous works \cite{HTG17,HT19,THSG20}, our initial objective has been to determine approximate expressions for the rheological properties of the molecular gas by using Grad's moment method \cite{G49}. Specifically, we have computed the collisional moment $\mathsf{A}_g$ defined in Eq.\ \eqref{3.7} through Grad's distribution \eqref{3.9}. The knowledge of this collisional moment enables the derivation of the explicit forms for the non-Newtonian shear viscosity $\eta_g$ and the viscometric function $\Psi^*_g$ of the gas phase \cite{G02,SGD04}. The anticipated shear-thinning effect in the shear viscosity is observed \cite{GS03}, a phenomenon corroborated by the excellent agreement between the results obtained from Grad's method and DSMC simulations.

Following the characterization of the molecular gas, we have also computed by means of Grad's moment method the collisional moments $\mathsf{B}$ and $\mathsf{C}$ of the granular gas defined in Eqs.\ \eqref{4.8} and \eqref{4.9}, respectively. After performing algebraic manipulations, we have expressed the (reduced) granular temperature $\chi$ and the (reduced) relevant components of the pressure tensor $P^*_{ij}$ in terms of the parameter space governing the problem. This parameter space includes the  coefficient of normal restitution $\alpha$, the (reduced) shear rate $a^*$, the (reduced) background temperature $T_g^*$, the volume fraction $\phi$, and the diameter $\sigma/\sigma_g$ and mass $m/m_g$ ratios. Consequently, a noteworthy contribution of our findings lies in providing an explicit dependence of the rheological properties on the above ratios, especially in the case of the mass ratio $m/m_g$.

The results derived here for arbitrary values of the mass ratio show no new surprises relative to the earlier works for gas-solid flows \cite{HTG17,GGG19}: the flow curve for the non-Newtonian viscosity $\eta^*$ exhibits an $S$-shape, and hence, DST is present. Nonetheless, from a more quantitative level, the results obtained in this work illustrate the amplifying influence of the DST effect with increasing the mass ratio $m/m_g$. Notably, when the ratio $m/m_g$ reaches a sufficiently large value, the conclusions derived in this study agree with those previously established in the context of the Brownian limit from the Fokker--Planck operator \cite{HTG17,GGG19}. This agreement serves to validate the application of the conventional Langevin equation in the analysis of non-Newtonian transport within dilute granular suspensions, even when subjected to high shear rates \cite{PZ23}. Another remarkable result lies in the mass ratio values for which agreement is reached with the coarse-grained results.  In contrast to the situation observed in the context of the Navier--Stokes transport coefficients, where convergence to the Brownian limit was attained for relatively small mass ratios ($m/m_g\approx 50$) \cite{GG22a}, the rheological scenario requires mass ratios exceeding $10^6$ to observe this collapse. This emphasizes the intricacy of the collisional model, not only in its capacity to elucidate the Fokker--Planck model as a limiting case derived from purely kinetic principles but also for its ability for representing a diverse spectrum of systems whose rheology may be not fully captured by the existing effective models.

Another noteworthy observation related to the variability of the mass ratio $m/m_g$ lies in the ability to completely modify the rheological behavior of the suspension. This is manifested in a transition from shear thinning to thickening for a critical mass ratio when we consider sufficiently low values of $a^*$ as showed in Figs.\ \ref{fig5}--\ref{fig5b}.

Furthermore, the results depicted in Figs.\ \ref{fig3}, \ref{fig4}, and \ref{fig5c} highlight the commendable efficacy of both Grad's method and DSMC simulations in capturing the shear-rate dependence of the rheological properties. Remarkably, a perfect agreement emerges among the results obtained from Grad's method, DSMC, and MD simulations in the Brownian limit. Despite this agreement, slight disparities manifest in a limited region near the transition point, where MD simulation data indicate a more pronounced transition than the Boltzmann results suggest. This minor disparity is attributed primarily to the constraints of the Boltzmann equation in incorporating small density corrections to $\eta^*$ in the vicinity of this transition point.

Although several mechanisms have been proposed in the literature (most of them related to to structural properties and mutual friction between grains \cite{BJ14, BNPDM22}), the knowledge of the non-Newtonian transport properties in granular suspensions is still a challenging open problem. What is interesting in the present work is the existence of this
shear thickening in a structurally simple system (low-density granular gas immersed in a dilute molecular gas). In this case, the non-Newtonian
properties of the granular gas are associated with both the behavior of the granular suspension in far from
equilibrium situations as well as the impact of the bath (molecular interstitial gas) on the dynamics properties of the solid particles.


Taking advantage of the excellent agreement observed between MD and DSMC computer simulations, we can exploit the latter to explore new situations in which to validate Grad's results. This is first done in Fig. \ref{fig5c}, where small mass ratios are considered to ensure comparable evolution times for the granular and molecular gases. We find a good agreement between theory and DSMC data, validating Grad's results for finite mass ratios. In a second contribution, Fig.\ \ref{fig6} presents the rheological properties plotted against shear rate in the Brownian limit for various coefficients of restitution. Notably, we observe again a good agreement between theoretical predictions and simulation data, highlighting the robustness of Grad's results even under conditions with moderate inelasticities. On the other hand, as expected, the influence of inelasticity manifests in a dampening of the DST transition. Although this dissipative effect is evident in the viscometric function $\Psi^*$ as well, its impact is comparatively less pronounced.

As an interesting work, we plan to extend the current findings to finite densities by considering the Enskog kinetic equation. Within this context, an intriguing question arises: does the variability of the mass ratio influence the transition from DST to CST as density increases, as observed in the Fokker--Planck model? Exploration along this avenue is planned for future project.

\acknowledgments

We are grateful to Dr. A. Zaccone for a critical reading of the manuscript. We acknowledge financial support from Grant PID2020-112936GB-I00 funded by MCIN/AEI/10.13039/501100011033, and  from Grant IB20079 funded by Junta de Extremadura (Spain) and by ERDF A way of making Europe. 

\bibliography{Brownian}

\begin{thebibliography}{71}%
\makeatletter
\providecommand \@ifxundefined [1]{%
 \@ifx{#1\undefined}
}%
\providecommand \@ifnum [1]{%
 \ifnum #1\expandafter \@firstoftwo
 \else \expandafter \@secondoftwo
 \fi
}%
\providecommand \@ifx [1]{%
 \ifx #1\expandafter \@firstoftwo
 \else \expandafter \@secondoftwo
 \fi
}%
\providecommand \natexlab [1]{#1}%
\providecommand \enquote  [1]{``#1''}%
\providecommand \bibnamefont  [1]{#1}%
\providecommand \bibfnamefont [1]{#1}%
\providecommand \citenamefont [1]{#1}%
\providecommand \href@noop [0]{\@secondoftwo}%
\providecommand \href [0]{\begingroup \@sanitize@url \@href}%
\providecommand \@href[1]{\@@startlink{#1}\@@href}%
\providecommand \@@href[1]{\endgroup#1\@@endlink}%
\providecommand \@sanitize@url [0]{\catcode `\\12\catcode `\$12\catcode
  `\&12\catcode `\#12\catcode `\^12\catcode `\_12\catcode `\%12\relax}%
\providecommand \@@startlink[1]{}%
\providecommand \@@endlink[0]{}%
\providecommand \url  [0]{\begingroup\@sanitize@url \@url }%
\providecommand \@url [1]{\endgroup\@href {#1}{\urlprefix }}%
\providecommand \urlprefix  [0]{URL }%
\providecommand \Eprint [0]{\href }%
\providecommand \doibase [0]{http://dx.doi.org/}%
\providecommand \selectlanguage [0]{\@gobble}%
\providecommand \bibinfo  [0]{\@secondoftwo}%
\providecommand \bibfield  [0]{\@secondoftwo}%
\providecommand \translation [1]{[#1]}%
\providecommand \BibitemOpen [0]{}%
\providecommand \bibitemStop [0]{}%
\providecommand \bibitemNoStop [0]{.\EOS\space}%
\providecommand \EOS [0]{\spacefactor3000\relax}%
\providecommand \BibitemShut  [1]{\csname bibitem#1\endcsname}%
\let\auto@bib@innerbib\@empty
\bibitem [{\citenamefont {Subramaniam}(2020)}]{S20}%
  \BibitemOpen
  \bibfield  {author} {\bibinfo {author} {\bibfnamefont {S.}~\bibnamefont
  {Subramaniam}},\ }\bibfield  {title} {\enquote {\bibinfo {title} {Multiphase
  flows: {R}ich physics, challenging theory, and big simualtions},}\
  }\href@noop {} {\bibfield  {journal} {\bibinfo  {journal} {Phys. Rev.
  Fluids}\ }\textbf {\bibinfo {volume} {5}},\ \bibinfo {pages} {110520}
  (\bibinfo {year} {2020})}\BibitemShut {NoStop}%
\bibitem [{\citenamefont {Koch}(1990)}]{K90}%
  \BibitemOpen
  \bibfield  {author} {\bibinfo {author} {\bibfnamefont {D.~L.}\ \bibnamefont
  {Koch}},\ }\bibfield  {title} {\enquote {\bibinfo {title} {Kinetic theory for
  a monodisperse gas-solid suspension},}\ }\href@noop {} {\bibfield  {journal}
  {\bibinfo  {journal} {Phys. Fluids A}\ }\textbf {\bibinfo {volume} {2}},\
  \bibinfo {pages} {1711--1722} (\bibinfo {year} {1990})}\BibitemShut {NoStop}%
\bibitem [{\citenamefont {Gidaspow}(1994)}]{G94}%
  \BibitemOpen
  \bibfield  {author} {\bibinfo {author} {\bibfnamefont {D.}~\bibnamefont
  {Gidaspow}},\ }\href@noop {} {\emph {\bibinfo {title} {Multiphase Flow and
  Fluidization}}}\ (\bibinfo  {publisher} {Academic Press},\ \bibinfo {year}
  {1994})\BibitemShut {NoStop}%
\bibitem [{\citenamefont {Jackson}(2000)}]{J00}%
  \BibitemOpen
  \bibfield  {author} {\bibinfo {author} {\bibfnamefont {R.}~\bibnamefont
  {Jackson}},\ }\href@noop {} {\emph {\bibinfo {title} {The {D}ynamics of
  {F}luidized {P}articles}}}\ (\bibinfo  {publisher} {Cambridge University
  Press, New York},\ \bibinfo {year} {2000})\BibitemShut {NoStop}%
\bibitem [{\citenamefont {Fullmer}\ and\ \citenamefont {Hrenya}(2017)}]{FH17}%
  \BibitemOpen
  \bibfield  {author} {\bibinfo {author} {\bibfnamefont {W.~D.}\ \bibnamefont
  {Fullmer}}\ and\ \bibinfo {author} {\bibfnamefont {C.~M.}\ \bibnamefont
  {Hrenya}},\ }\bibfield  {title} {\enquote {\bibinfo {title} {The clustering
  instability in rapid granular and gas-solid flows},}\ }\href@noop {}
  {\bibfield  {journal} {\bibinfo  {journal} {Annu. Rev. Fluid Mech.}\ }\textbf
  {\bibinfo {volume} {49}},\ \bibinfo {pages} {485--510} (\bibinfo {year}
  {2017})}\BibitemShut {NoStop}%
\bibitem [{\citenamefont {Zhao}\ and\ \citenamefont {Wang}(2021)}]{ZW21}%
  \BibitemOpen
  \bibfield  {author} {\bibinfo {author} {\bibfnamefont {B.}~\bibnamefont
  {Zhao}}\ and\ \bibinfo {author} {\bibfnamefont {J.}~\bibnamefont {Wang}},\
  }\bibfield  {title} {\enquote {\bibinfo {title} {Kinetic theory of
  polydisperse gas-solid flow: {N}avier-stokes transport coefficients},}\
  }\href@noop {} {\bibfield  {journal} {\bibinfo  {journal} {Phys. Fluids}\
  }\textbf {\bibinfo {volume} {33}},\ \bibinfo {pages} {103322} (\bibinfo
  {year} {2021})}\BibitemShut {NoStop}%
\bibitem [{\citenamefont {Chassagne}\ \emph {et~al.}(2023)\citenamefont
  {Chassagne}, \citenamefont {Bonamy},\ and\ \citenamefont {Chauchat}}]{CBC23}%
  \BibitemOpen
  \bibfield  {author} {\bibinfo {author} {\bibfnamefont {R.}~\bibnamefont
  {Chassagne}}, \bibinfo {author} {\bibfnamefont {C.}~\bibnamefont {Bonamy}}, \
  and\ \bibinfo {author} {\bibfnamefont {J.}~\bibnamefont {Chauchat}},\
  }\bibfield  {title} {\enquote {\bibinfo {title} {A frictional-collisional
  model for bedload transport based on kinetic theory of granular flows:
  discrete and continuum approaches},}\ }\href@noop {} {\bibfield  {journal}
  {\bibinfo  {journal} {J. Fluid Mech.}\ }\textbf {\bibinfo {volume} {964}},\
  \bibinfo {pages} {A27} (\bibinfo {year} {2023})}\BibitemShut {NoStop}%
\bibitem [{\citenamefont {Tsao}\ and\ \citenamefont {Koch}(1995)}]{TK95}%
  \BibitemOpen
  \bibfield  {author} {\bibinfo {author} {\bibfnamefont {H.-K.}\ \bibnamefont
  {Tsao}}\ and\ \bibinfo {author} {\bibfnamefont {D.~L.}\ \bibnamefont
  {Koch}},\ }\bibfield  {title} {\enquote {\bibinfo {title} {Simple shear flows
  of dilute gas-solid suspensions},}\ }\href@noop {} {\bibfield  {journal}
  {\bibinfo  {journal} {J. Fluid Mech.}\ }\textbf {\bibinfo {volume} {296}},\
  \bibinfo {pages} {211--245} (\bibinfo {year} {1995})}\BibitemShut {NoStop}%
\bibitem [{\citenamefont {Sangani}\ \emph {et~al.}(1996)\citenamefont
  {Sangani}, \citenamefont {Mo}, \citenamefont {Tsao},\ and\ \citenamefont
  {Koch}}]{SMTK96}%
  \BibitemOpen
  \bibfield  {author} {\bibinfo {author} {\bibfnamefont {A.~S.}\ \bibnamefont
  {Sangani}}, \bibinfo {author} {\bibfnamefont {G.}~\bibnamefont {Mo}},
  \bibinfo {author} {\bibfnamefont {H.-K.}\ \bibnamefont {Tsao}}, \ and\
  \bibinfo {author} {\bibfnamefont {D.~L.}\ \bibnamefont {Koch}},\ }\bibfield
  {title} {\enquote {\bibinfo {title} {Simple shear flows of dense gas-solid
  suspensions at finite \textsc{S}tokes numbers},}\ }\href@noop {} {\bibfield
  {journal} {\bibinfo  {journal} {J. Fluid Mech.}\ }\textbf {\bibinfo {volume}
  {313}},\ \bibinfo {pages} {309--341} (\bibinfo {year} {1996})}\BibitemShut
  {NoStop}%
\bibitem [{\citenamefont {Wylie}\ \emph
  {et~al.}(2009{\natexlab{a}})\citenamefont {Wylie}, \citenamefont {Zhang},
  \citenamefont {Li},\ and\ \citenamefont {Hengyi}}]{WZLH09}%
  \BibitemOpen
  \bibfield  {author} {\bibinfo {author} {\bibfnamefont {J.~J.}\ \bibnamefont
  {Wylie}}, \bibinfo {author} {\bibfnamefont {Q.}~\bibnamefont {Zhang}},
  \bibinfo {author} {\bibfnamefont {Y.}~\bibnamefont {Li}}, \ and\ \bibinfo
  {author} {\bibfnamefont {X.}~\bibnamefont {Hengyi}},\ }\bibfield  {title}
  {\enquote {\bibinfo {title} {Driven inelastic-particle systems with drag},}\
  }\href@noop {} {\bibfield  {journal} {\bibinfo  {journal} {Phys. Rev. E}\
  }\textbf {\bibinfo {volume} {79}},\ \bibinfo {pages} {031301} (\bibinfo
  {year} {2009}{\natexlab{a}})}\BibitemShut {NoStop}%
\bibitem [{\citenamefont {Heussinger}(2013)}]{H13}%
  \BibitemOpen
  \bibfield  {author} {\bibinfo {author} {\bibfnamefont {C.}~\bibnamefont
  {Heussinger}},\ }\bibfield  {title} {\enquote {\bibinfo {title} {Shear
  thickening in granular suspensions: {I}nterparticle friction and dynamically
  correlated clusters},}\ }\href@noop {} {\bibfield  {journal} {\bibinfo
  {journal} {Phys. Rev. E}\ }\textbf {\bibinfo {volume} {88}},\ \bibinfo
  {pages} {050201 (R)} (\bibinfo {year} {2013})}\BibitemShut {NoStop}%
\bibitem [{\citenamefont {Chamorro}\ \emph {et~al.}(2015)\citenamefont
  {Chamorro}, \citenamefont {Vega~Reyes},\ and\ \citenamefont
  {Garz\'o}}]{ChVG15}%
  \BibitemOpen
  \bibfield  {author} {\bibinfo {author} {\bibfnamefont {M.~G.}\ \bibnamefont
  {Chamorro}}, \bibinfo {author} {\bibfnamefont {F.}~\bibnamefont
  {Vega~Reyes}}, \ and\ \bibinfo {author} {\bibfnamefont {V.}~\bibnamefont
  {Garz\'o}},\ }\bibfield  {title} {\enquote {\bibinfo {title} {Non-{N}ewtonian
  hydrodynamics for a dilute granular suspension under uniform shear flow},}\
  }\href@noop {} {\bibfield  {journal} {\bibinfo  {journal} {Phys. Rev. E}\
  }\textbf {\bibinfo {volume} {92}},\ \bibinfo {pages} {{052}{205}} (\bibinfo
  {year} {2015})}\BibitemShut {NoStop}%
\bibitem [{\citenamefont {Saha}\ and\ \citenamefont {Alam}(2017)}]{SA17}%
  \BibitemOpen
  \bibfield  {author} {\bibinfo {author} {\bibfnamefont {S.}~\bibnamefont
  {Saha}}\ and\ \bibinfo {author} {\bibfnamefont {M.}~\bibnamefont {Alam}},\
  }\bibfield  {title} {\enquote {\bibinfo {title} {Revisiting ignited-quenched
  transition and the non-{N}ewtonian rheology of a sheared dilute gas-solid
  suspension},}\ }\href@noop {} {\bibfield  {journal} {\bibinfo  {journal} {J.
  Fluid Mech.}\ }\textbf {\bibinfo {volume} {833}},\ \bibinfo {pages}
  {206--246} (\bibinfo {year} {2017})}\BibitemShut {NoStop}%
\bibitem [{\citenamefont {Alam}\ \emph {et~al.}(2019)\citenamefont {Alam},
  \citenamefont {Saha},\ and\ \citenamefont {Gupta}}]{ASG19}%
  \BibitemOpen
  \bibfield  {author} {\bibinfo {author} {\bibfnamefont {M.}~\bibnamefont
  {Alam}}, \bibinfo {author} {\bibfnamefont {S.}~\bibnamefont {Saha}}, \ and\
  \bibinfo {author} {\bibfnamefont {R.}~\bibnamefont {Gupta}},\ }\bibfield
  {title} {\enquote {\bibinfo {title} {Unified theory for a sheared gas-solid
  suspension: from rapid granular suspension to its small-{S}tokes-number
  limit},}\ }\href@noop {} {\bibfield  {journal} {\bibinfo  {journal} {J. Fluid
  Mech.}\ }\textbf {\bibinfo {volume} {870}},\ \bibinfo {pages} {1175--1193}
  (\bibinfo {year} {2019})}\BibitemShut {NoStop}%
\bibitem [{\citenamefont {Saha}\ and\ \citenamefont {Alam}(2020)}]{SA20}%
  \BibitemOpen
  \bibfield  {author} {\bibinfo {author} {\bibfnamefont {S.}~\bibnamefont
  {Saha}}\ and\ \bibinfo {author} {\bibfnamefont {M.}~\bibnamefont {Alam}},\
  }\bibfield  {title} {\enquote {\bibinfo {title} {Burnett-order constitutive
  relations, second moment anisotropy and co-existing states in sheared dense
  gas-solid suspensions},}\ }\href@noop {} {\bibfield  {journal} {\bibinfo
  {journal} {J. Fluid Mech.}\ }\textbf {\bibinfo {volume} {887}},\ \bibinfo
  {pages} {A9} (\bibinfo {year} {2020})}\BibitemShut {NoStop}%
\bibitem [{\citenamefont {Garz\'o}\ \emph {et~al.}(2012)\citenamefont
  {Garz\'o}, \citenamefont {Tenneti}, \citenamefont {Subramaniam},\ and\
  \citenamefont {Hrenya}}]{GTSH12}%
  \BibitemOpen
  \bibfield  {author} {\bibinfo {author} {\bibfnamefont {V.}~\bibnamefont
  {Garz\'o}}, \bibinfo {author} {\bibfnamefont {S.}~\bibnamefont {Tenneti}},
  \bibinfo {author} {\bibfnamefont {S.}~\bibnamefont {Subramaniam}}, \ and\
  \bibinfo {author} {\bibfnamefont {C.~M.}\ \bibnamefont {Hrenya}},\ }\bibfield
   {title} {\enquote {\bibinfo {title} {Enskog kinetic theory for monodisperse
  gas-solid flows},}\ }\href@noop {} {\bibfield  {journal} {\bibinfo  {journal}
  {J. Fluid Mech.}\ }\textbf {\bibinfo {volume} {712}},\ \bibinfo {pages}
  {129--168} (\bibinfo {year} {2012})}\BibitemShut {NoStop}%
\bibitem [{\citenamefont {Chapman}\ and\ \citenamefont {Cowling}(1970)}]{CC70}%
  \BibitemOpen
  \bibfield  {author} {\bibinfo {author} {\bibfnamefont {S.}~\bibnamefont
  {Chapman}}\ and\ \bibinfo {author} {\bibfnamefont {T.~G.}\ \bibnamefont
  {Cowling}},\ }\href@noop {} {\emph {\bibinfo {title} {The Mathematical Theory
  of Nonuniform Gases}}}\ (\bibinfo  {publisher} {Cambridge University Press,
  Cambridge},\ \bibinfo {year} {1970})\BibitemShut {NoStop}%
\bibitem [{\citenamefont {Garz\'o}\ \emph
  {et~al.}(2013{\natexlab{a}})\citenamefont {Garz\'o}, \citenamefont
  {Chamorro},\ and\ \citenamefont {Vega~Reyes}}]{GChV13a}%
  \BibitemOpen
  \bibfield  {author} {\bibinfo {author} {\bibfnamefont {V.}~\bibnamefont
  {Garz\'o}}, \bibinfo {author} {\bibfnamefont {M.~G.}\ \bibnamefont
  {Chamorro}}, \ and\ \bibinfo {author} {\bibfnamefont {F.}~\bibnamefont
  {Vega~Reyes}},\ }\bibfield  {title} {\enquote {\bibinfo {title} {Transport
  properties for driven granular fluids in situations close to homogeneous
  steady states},}\ }\href@noop {} {\bibfield  {journal} {\bibinfo  {journal}
  {Phys. Rev. E}\ }\textbf {\bibinfo {volume} {87}},\ \bibinfo {pages}
  {{032}{201}} (\bibinfo {year} {2013}{\natexlab{a}})}\BibitemShut {NoStop}%
\bibitem [{\citenamefont {Garz\'o}\ \emph
  {et~al.}(2013{\natexlab{b}})\citenamefont {Garz\'o}, \citenamefont
  {Chamorro},\ and\ \citenamefont {Vega~Reyes}}]{GChV13b}%
  \BibitemOpen
  \bibfield  {author} {\bibinfo {author} {\bibfnamefont {V.}~\bibnamefont
  {Garz\'o}}, \bibinfo {author} {\bibfnamefont {M.~G.}\ \bibnamefont
  {Chamorro}}, \ and\ \bibinfo {author} {\bibfnamefont {F.}~\bibnamefont
  {Vega~Reyes}},\ }\bibfield  {title} {\enquote {\bibinfo {title} {Erratum:
  {T}ransport properties for driven granular fluids in situations close to
  homogeneous steady states},}\ }\href@noop {} {\bibfield  {journal} {\bibinfo
  {journal} {Phys. Rev. E}\ }\textbf {\bibinfo {volume} {87}},\ \bibinfo
  {pages} {{059}{906}} (\bibinfo {year} {2013}{\natexlab{b}})}\BibitemShut
  {NoStop}%
\bibitem [{\citenamefont {G\'omez~Gonz\'alez}\ and\ \citenamefont
  {Garz\'o}(2023)}]{GG23}%
  \BibitemOpen
  \bibfield  {author} {\bibinfo {author} {\bibfnamefont {R.}~\bibnamefont
  {G\'omez~Gonz\'alez}}\ and\ \bibinfo {author} {\bibfnamefont
  {V.}~\bibnamefont {Garz\'o}},\ }\bibfield  {title} {\enquote {\bibinfo
  {title} {Tracer diffusion coefficients in a moderately dense granular
  suspension: {S}tability analysis and thermal diffusion segregation},}\
  }\href@noop {} {\bibfield  {journal} {\bibinfo  {journal} {Phys. Fluids}\
  }\textbf {\bibinfo {volume} {35}},\ \bibinfo {pages} {083318} (\bibinfo
  {year} {2023})}\BibitemShut {NoStop}%
\bibitem [{\citenamefont {Seto}\ \emph {et~al.}(2013)\citenamefont {Seto},
  \citenamefont {Mari}, \citenamefont {Morris},\ and\ \citenamefont
  {Denn}}]{SMMD13}%
  \BibitemOpen
  \bibfield  {author} {\bibinfo {author} {\bibfnamefont {R.}~\bibnamefont
  {Seto}}, \bibinfo {author} {\bibfnamefont {R.}~\bibnamefont {Mari}}, \bibinfo
  {author} {\bibfnamefont {J.~F.}\ \bibnamefont {Morris}}, \ and\ \bibinfo
  {author} {\bibfnamefont {M.~M.}\ \bibnamefont {Denn}},\ }\bibfield  {title}
  {\enquote {\bibinfo {title} {Discontinuous shear thickening of frictional
  hard-sphere suspensions},}\ }\href@noop {} {\bibfield  {journal} {\bibinfo
  {journal} {Phys. Rev. Lett.}\ }\textbf {\bibinfo {volume} {111}},\ \bibinfo
  {pages} {{218}{301}} (\bibinfo {year} {2013})}\BibitemShut {NoStop}%
\bibitem [{\citenamefont {Kawasaki}\ \emph {et~al.}(2014)\citenamefont
  {Kawasaki}, \citenamefont {Ikeda},\ and\ \citenamefont {Berthier}}]{KIB14}%
  \BibitemOpen
  \bibfield  {author} {\bibinfo {author} {\bibfnamefont {T.}~\bibnamefont
  {Kawasaki}}, \bibinfo {author} {\bibfnamefont {A.}~\bibnamefont {Ikeda}}, \
  and\ \bibinfo {author} {\bibfnamefont {L.}~\bibnamefont {Berthier}},\
  }\bibfield  {title} {\enquote {\bibinfo {title} {Thinning or thickening?
  {M}ultiple rheological regimes in dense suspensions of soft particles},}\
  }\href@noop {} {\bibfield  {journal} {\bibinfo  {journal} {EPL}\ }\textbf
  {\bibinfo {volume} {107}},\ \bibinfo {pages} {28009} (\bibinfo {year}
  {2014})}\BibitemShut {NoStop}%
\bibitem [{\citenamefont {Hayakawa}\ \emph {et~al.}(2017)\citenamefont
  {Hayakawa}, \citenamefont {Takada},\ and\ \citenamefont {Garz\'o}}]{HTG17}%
  \BibitemOpen
  \bibfield  {author} {\bibinfo {author} {\bibfnamefont {H.}~\bibnamefont
  {Hayakawa}}, \bibinfo {author} {\bibfnamefont {S.}~\bibnamefont {Takada}}, \
  and\ \bibinfo {author} {\bibfnamefont {V.}~\bibnamefont {Garz\'o}},\
  }\bibfield  {title} {\enquote {\bibinfo {title} {Kinetic theory of shear
  thickening for a moderately dense gas-solid suspension: {F}rom discontinuous
  thickening to continuous thickening},}\ }\href@noop {} {\bibfield  {journal}
  {\bibinfo  {journal} {Phys. Rev. E}\ }\textbf {\bibinfo {volume} {96}},\
  \bibinfo {pages} {042903} (\bibinfo {year} {2017})}\BibitemShut {NoStop}%
\bibitem [{\citenamefont {Hayakawa}\ \emph {et~al.}(2020)\citenamefont
  {Hayakawa}, \citenamefont {Takada},\ and\ \citenamefont {Garz\'o}}]{HTG20}%
  \BibitemOpen
  \bibfield  {author} {\bibinfo {author} {\bibfnamefont {H.}~\bibnamefont
  {Hayakawa}}, \bibinfo {author} {\bibfnamefont {S.}~\bibnamefont {Takada}}, \
  and\ \bibinfo {author} {\bibfnamefont {V.}~\bibnamefont {Garz\'o}},\
  }\bibfield  {title} {\enquote {\bibinfo {title} {Erratum: Kinetic theory of
  shear thickening for a moderately dense gas-solid suspension: from
  discontinuous thickening to continuous thickening},}\ }\href@noop {}
  {\bibfield  {journal} {\bibinfo  {journal} {Phys. Rev. E}\ }\textbf {\bibinfo
  {volume} {101}},\ \bibinfo {pages} {069904(E)} (\bibinfo {year}
  {2020})}\BibitemShut {NoStop}%
\bibitem [{\citenamefont {Hayakawa}\ and\ \citenamefont {Takada}(2019)}]{HT19}%
  \BibitemOpen
  \bibfield  {author} {\bibinfo {author} {\bibfnamefont {H.}~\bibnamefont
  {Hayakawa}}\ and\ \bibinfo {author} {\bibfnamefont {S.}~\bibnamefont
  {Takada}},\ }\bibfield  {title} {\enquote {\bibinfo {title} {Kinetic theory
  of discontinuous rheological phase transition for a dilute inertial
  suspension},}\ }\href@noop {} {\bibfield  {journal} {\bibinfo  {journal}
  {Prog. Theor. Exp. Phys}\ }\textbf {\bibinfo {volume} {083J01}} (\bibinfo
  {year} {2019})}\BibitemShut {NoStop}%
\bibitem [{\citenamefont {Takada}\ \emph {et~al.}(2020)\citenamefont {Takada},
  \citenamefont {Hayakawa}, \citenamefont {Santos},\ and\ \citenamefont
  {Garz\'o}}]{THSG20}%
  \BibitemOpen
  \bibfield  {author} {\bibinfo {author} {\bibfnamefont {S.}~\bibnamefont
  {Takada}}, \bibinfo {author} {\bibfnamefont {H.}~\bibnamefont {Hayakawa}},
  \bibinfo {author} {\bibfnamefont {A.}~\bibnamefont {Santos}}, \ and\ \bibinfo
  {author} {\bibfnamefont {V.}~\bibnamefont {Garz\'o}},\ }\bibfield  {title}
  {\enquote {\bibinfo {title} {Enskog kinetic theory of rheology for a
  moderately dense inertial suspension},}\ }\href@noop {} {\bibfield  {journal}
  {\bibinfo  {journal} {Phys. Rev. E}\ }\textbf {\bibinfo {volume} {102}},\
  \bibinfo {pages} {022907} (\bibinfo {year} {2020})}\BibitemShut {NoStop}%
\bibitem [{\citenamefont {Takada}\ \emph {et~al.}(2023)\citenamefont {Takada},
  \citenamefont {Hayakawa},\ and\ \citenamefont {Garz\'o}}]{THG23}%
  \BibitemOpen
  \bibfield  {author} {\bibinfo {author} {\bibfnamefont {S.}~\bibnamefont
  {Takada}}, \bibinfo {author} {\bibfnamefont {H.}~\bibnamefont {Hayakawa}}, \
  and\ \bibinfo {author} {\bibfnamefont {V.}~\bibnamefont {Garz\'o}},\
  }\bibfield  {title} {\enquote {\bibinfo {title} {Rheology of a dilute binary
  mixture of inertial suspension under simple shear flow},}\ }\href@noop {}
  {\bibfield  {journal} {\bibinfo  {journal} {Prog. Theor. Exp. Phys.}\
  }\textbf {\bibinfo {volume} {113J01}} (\bibinfo {year} {2023})}\BibitemShut
  {NoStop}%
\bibitem [{\citenamefont {Capecelatro}\ and\ \citenamefont
  {Desjardins}(2013)}]{CD13}%
  \BibitemOpen
  \bibfield  {author} {\bibinfo {author} {\bibfnamefont {J.}~\bibnamefont
  {Capecelatro}}\ and\ \bibinfo {author} {\bibfnamefont {O.}~\bibnamefont
  {Desjardins}},\ }\bibfield  {title} {\enquote {\bibinfo {title} {An
  {E}uler--{L}agrange strategy for simulating particle-laden flows},}\
  }\href@noop {} {\bibfield  {journal} {\bibinfo  {journal} {J. Comput. Phys.}\
  }\textbf {\bibinfo {volume} {238}},\ \bibinfo {pages} {1} (\bibinfo {year}
  {2013})}\BibitemShut {NoStop}%
\bibitem [{\citenamefont {Capecelatro}\ \emph {et~al.}(2015)\citenamefont
  {Capecelatro}, \citenamefont {Desjardins},\ and\ \citenamefont
  {Fox}}]{CDF15}%
  \BibitemOpen
  \bibfield  {author} {\bibinfo {author} {\bibfnamefont {J.}~\bibnamefont
  {Capecelatro}}, \bibinfo {author} {\bibfnamefont {O.}~\bibnamefont
  {Desjardins}}, \ and\ \bibinfo {author} {\bibfnamefont {R.~O.}\ \bibnamefont
  {Fox}},\ }\bibfield  {title} {\enquote {\bibinfo {title} {On fluid-particle
  dynamics in fully developed cluster-induced turbulence},}\ }\href@noop {}
  {\bibfield  {journal} {\bibinfo  {journal} {J. Fluid Mech.}\ }\textbf
  {\bibinfo {volume} {780}},\ \bibinfo {pages} {578} (\bibinfo {year}
  {2015})}\BibitemShut {NoStop}%
\bibitem [{\citenamefont {Garz\'o}\ \emph {et~al.}(2016)\citenamefont
  {Garz\'o}, \citenamefont {Fullmer}, \citenamefont {Hrenya},\ and\
  \citenamefont {Yin}}]{GFHY16}%
  \BibitemOpen
  \bibfield  {author} {\bibinfo {author} {\bibfnamefont {V.}~\bibnamefont
  {Garz\'o}}, \bibinfo {author} {\bibfnamefont {W.~D.}\ \bibnamefont
  {Fullmer}}, \bibinfo {author} {\bibfnamefont {C.~M.}\ \bibnamefont {Hrenya}},
  \ and\ \bibinfo {author} {\bibfnamefont {X.}~\bibnamefont {Yin}},\ }\bibfield
   {title} {\enquote {\bibinfo {title} {Transport coefficients of solid
  particles immersed in a viscous gas},}\ }\href@noop {} {\bibfield  {journal}
  {\bibinfo  {journal} {Phys. Rev. E}\ }\textbf {\bibinfo {volume} {93}},\
  \bibinfo {pages} {012905} (\bibinfo {year} {2016})}\BibitemShut {NoStop}%
\bibitem [{\citenamefont {Fullmer}\ \emph {et~al.}(2017)\citenamefont
  {Fullmer}, \citenamefont {Liu}, \citenamefont {Yin},\ and\ \citenamefont
  {Hrenya}}]{FLYH17}%
  \BibitemOpen
  \bibfield  {author} {\bibinfo {author} {\bibfnamefont {W.~D.}\ \bibnamefont
  {Fullmer}}, \bibinfo {author} {\bibfnamefont {G.}~\bibnamefont {Liu}},
  \bibinfo {author} {\bibfnamefont {X.}~\bibnamefont {Yin}}, \ and\ \bibinfo
  {author} {\bibfnamefont {C.~M.}\ \bibnamefont {Hrenya}},\ }\bibfield  {title}
  {\enquote {\bibinfo {title} {Clustering instabilities in sedimenting
  fluid-solid systems: critical assessment of kinetic-theory-based predictions
  using the direct numerical simulation data},}\ }\href@noop {} {\bibfield
  {journal} {\bibinfo  {journal} {J. Fluid Mech.}\ }\textbf {\bibinfo {volume}
  {823}},\ \bibinfo {pages} {433} (\bibinfo {year} {2017})}\BibitemShut
  {NoStop}%
\bibitem [{\citenamefont {Pelargonio}\ and\ \citenamefont
  {Zaccone}(2023)}]{PZ23}%
  \BibitemOpen
  \bibfield  {author} {\bibinfo {author} {\bibfnamefont {S.}~\bibnamefont
  {Pelargonio}}\ and\ \bibinfo {author} {\bibfnamefont {A.}~\bibnamefont
  {Zaccone}},\ }\bibfield  {title} {\enquote {\bibinfo {title} {Generalized
  {L}angevin equation with shear flow and its fluctuation-dissipation theorems
  derived from a {C}aldeira-{L}eggett {H}amiltonian},}\ }\href@noop {}
  {\bibfield  {journal} {\bibinfo  {journal} {Phys. Rev. E}\ }\textbf {\bibinfo
  {volume} {107}},\ \bibinfo {pages} {064102} (\bibinfo {year}
  {2023})}\BibitemShut {NoStop}%
\bibitem [{\citenamefont {Biben}\ \emph {et~al.}(2002)\citenamefont {Biben},
  \citenamefont {Martin},\ and\ \citenamefont {Piasecki}}]{BMP02a}%
  \BibitemOpen
  \bibfield  {author} {\bibinfo {author} {\bibfnamefont {T.}~\bibnamefont
  {Biben}}, \bibinfo {author} {\bibfnamefont {Ph.A.}\ \bibnamefont {Martin}}, \
  and\ \bibinfo {author} {\bibfnamefont {J.}~\bibnamefont {Piasecki}},\
  }\bibfield  {title} {\enquote {\bibinfo {title} {Stationary state of
  thermostated inelastic hard spheres},}\ }\href@noop {} {\bibfield  {journal}
  {\bibinfo  {journal} {Physica A}\ }\textbf {\bibinfo {volume} {310}},\
  \bibinfo {pages} {308--324} (\bibinfo {year} {2002})}\BibitemShut {NoStop}%
\bibitem [{\citenamefont {G\'omez~Gonz\'alez}\ and\ \citenamefont
  {Garz\'o}(2022)}]{GG22a}%
  \BibitemOpen
  \bibfield  {author} {\bibinfo {author} {\bibfnamefont {R.}~\bibnamefont
  {G\'omez~Gonz\'alez}}\ and\ \bibinfo {author} {\bibfnamefont
  {V.}~\bibnamefont {Garz\'o}},\ }\bibfield  {title} {\enquote {\bibinfo
  {title} {Kinetic theory of granular particles immersed in a molecular gas},}\
  }\href@noop {} {\bibfield  {journal} {\bibinfo  {journal} {J. Fluid Mech.}\
  }\textbf {\bibinfo {volume} {943}},\ \bibinfo {pages} {A9} (\bibinfo {year}
  {2022})}\BibitemShut {NoStop}%
\bibitem [{\citenamefont {Fullmer}\ and\ \citenamefont {Hrenya}(2016)}]{FH16}%
  \BibitemOpen
  \bibfield  {author} {\bibinfo {author} {\bibfnamefont {W.~D.}\ \bibnamefont
  {Fullmer}}\ and\ \bibinfo {author} {\bibfnamefont {C.~M.}\ \bibnamefont
  {Hrenya}},\ }\bibfield  {title} {\enquote {\bibinfo {title} {Quantitative
  assessment of fine-grid kinetic-theory-based predictions},}\ }\href@noop {}
  {\bibfield  {journal} {\bibinfo  {journal} {AIChE}\ }\textbf {\bibinfo
  {volume} {62}},\ \bibinfo {pages} {11--17} (\bibinfo {year}
  {2016})}\BibitemShut {NoStop}%
\bibitem [{\citenamefont {G\'omez~Gonz\'alez}\ and\ \citenamefont
  {Garz\'o}(2019{\natexlab{a}})}]{GGG19a}%
  \BibitemOpen
  \bibfield  {author} {\bibinfo {author} {\bibfnamefont {R.}~\bibnamefont
  {G\'omez~Gonz\'alez}}\ and\ \bibinfo {author} {\bibfnamefont
  {V.}~\bibnamefont {Garz\'o}},\ }\bibfield  {title} {\enquote {\bibinfo
  {title} {Transport coefficients for granular suspensions at moderate
  densities},}\ }\href@noop {} {\bibfield  {journal} {\bibinfo  {journal} {J.
  Stat. Mech.}\ }\textbf {\bibinfo {volume} {093204}} (\bibinfo {year}
  {2019}{\natexlab{a}})}\BibitemShut {NoStop}%
\bibitem [{\citenamefont {Evans}\ and\ \citenamefont {Morriss}(1990)}]{EM90}%
  \BibitemOpen
  \bibfield  {author} {\bibinfo {author} {\bibfnamefont {D.~J.}\ \bibnamefont
  {Evans}}\ and\ \bibinfo {author} {\bibfnamefont {G.~P.}\ \bibnamefont
  {Morriss}},\ }\href@noop {} {\emph {\bibinfo {title} {Statistical Mechanics
  of Nonequilibrium Liquids}}}\ (\bibinfo  {publisher} {Academic Press,
  London},\ \bibinfo {year} {1990})\BibitemShut {NoStop}%
\bibitem [{\citenamefont {Garz\'o}\ and\ \citenamefont {Santos}(2003)}]{GS03}%
  \BibitemOpen
  \bibfield  {author} {\bibinfo {author} {\bibfnamefont {V.}~\bibnamefont
  {Garz\'o}}\ and\ \bibinfo {author} {\bibfnamefont {A.}~\bibnamefont
  {Santos}},\ }\href@noop {} {\emph {\bibinfo {title} {Kinetic Theory of Gases
  in Shear Flows. Nonlinear Transport}}}\ (\bibinfo  {publisher} {Springer,
  Netherlands},\ \bibinfo {year} {2003})\BibitemShut {NoStop}%
\bibitem [{\citenamefont {Dufty}\ \emph {et~al.}(1986)\citenamefont {Dufty},
  \citenamefont {Santos}, \citenamefont {Brey},\ and\ \citenamefont
  {Rodr\'{\i}guez}}]{DSBR86}%
  \BibitemOpen
  \bibfield  {author} {\bibinfo {author} {\bibfnamefont {J.~W.}\ \bibnamefont
  {Dufty}}, \bibinfo {author} {\bibfnamefont {A.}~\bibnamefont {Santos}},
  \bibinfo {author} {\bibfnamefont {J.~J.}\ \bibnamefont {Brey}}, \ and\
  \bibinfo {author} {\bibfnamefont {R.~F.}\ \bibnamefont {Rodr\'{\i}guez}},\
  }\bibfield  {title} {\enquote {\bibinfo {title} {Model for nonequilibrium
  computer simulation methods},}\ }\href@noop {} {\bibfield  {journal}
  {\bibinfo  {journal} {Phys. Rev. A}\ }\textbf {\bibinfo {volume} {33}},\
  \bibinfo {pages} {459--466} (\bibinfo {year} {1986})}\BibitemShut {NoStop}%
\bibitem [{\citenamefont {Grad}(1949)}]{G49}%
  \BibitemOpen
  \bibfield  {author} {\bibinfo {author} {\bibfnamefont {H.}~\bibnamefont
  {Grad}},\ }\bibfield  {title} {\enquote {\bibinfo {title} {On the kinetic
  theory of rarefied gases},}\ }\href@noop {} {\bibfield  {journal} {\bibinfo
  {journal} {Commun. Pure Appl. Math.}\ }\textbf {\bibinfo {volume} {2}},\
  \bibinfo {pages} {331--407} (\bibinfo {year} {1949})}\BibitemShut {NoStop}%
\bibitem [{\citenamefont {G\'omez~Gonz\'alez}\ and\ \citenamefont
  {Garz\'o}(2019{\natexlab{b}})}]{GGG19}%
  \BibitemOpen
  \bibfield  {author} {\bibinfo {author} {\bibfnamefont {R.}~\bibnamefont
  {G\'omez~Gonz\'alez}}\ and\ \bibinfo {author} {\bibfnamefont
  {V.}~\bibnamefont {Garz\'o}},\ }\bibfield  {title} {\enquote {\bibinfo
  {title} {Simple shear flow in granular suspensiones: {I}nelastic {M}axwell
  models and {B}{G}{K}-type kinetic model},}\ }\href@noop {} {\bibfield
  {journal} {\bibinfo  {journal} {J. Stat. Mech.}\ }\textbf {\bibinfo {volume}
  {013206}} (\bibinfo {year} {2019}{\natexlab{b}})}\BibitemShut {NoStop}%
\bibitem [{\citenamefont {Cercignani}(1988)}]{C88}%
  \BibitemOpen
  \bibfield  {author} {\bibinfo {author} {\bibfnamefont {C.}~\bibnamefont
  {Cercignani}},\ }\href@noop {} {\emph {\bibinfo {title} {The Boltzmann
  Equation and Its Applications}}}\ (\bibinfo  {publisher} {Springer--Verlag,
  New York},\ \bibinfo {year} {1988})\BibitemShut {NoStop}%
\bibitem [{\citenamefont {Garz\'o}(2019)}]{G19}%
  \BibitemOpen
  \bibfield  {author} {\bibinfo {author} {\bibfnamefont {V.}~\bibnamefont
  {Garz\'o}},\ }\href@noop {} {\emph {\bibinfo {title} {Granular Gaseous
  Flows}}}\ (\bibinfo  {publisher} {Springer Nature, Cham},\ \bibinfo {year}
  {2019})\BibitemShut {NoStop}%
\bibitem [{\citenamefont {R\'esibois}\ and\ \citenamefont
  {de~Leener}(1977)}]{RL77}%
  \BibitemOpen
  \bibfield  {author} {\bibinfo {author} {\bibfnamefont {P.}~\bibnamefont
  {R\'esibois}}\ and\ \bibinfo {author} {\bibfnamefont {M.}~\bibnamefont
  {de~Leener}},\ }\href@noop {} {\emph {\bibinfo {title} {Classical Kinetic
  Theory of Fluids}}}\ (\bibinfo  {publisher} {Wiley, New York},\ \bibinfo
  {year} {1977})\BibitemShut {NoStop}%
\bibitem [{\citenamefont {Garz\'o}(2002)}]{G02}%
  \BibitemOpen
  \bibfield  {author} {\bibinfo {author} {\bibfnamefont {V.}~\bibnamefont
  {Garz\'o}},\ }\bibfield  {title} {\enquote {\bibinfo {title} {Tracer
  diffusion in granular shear flows},}\ }\href@noop {} {\bibfield  {journal}
  {\bibinfo  {journal} {Phys. Rev. E}\ }\textbf {\bibinfo {volume} {66}},\
  \bibinfo {pages} {{021}{308}} (\bibinfo {year} {2002})}\BibitemShut {NoStop}%
\bibitem [{\citenamefont {Santos}\ \emph {et~al.}(2004)\citenamefont {Santos},
  \citenamefont {Garz\'o},\ and\ \citenamefont {Dufty}}]{SGD04}%
  \BibitemOpen
  \bibfield  {author} {\bibinfo {author} {\bibfnamefont {A.}~\bibnamefont
  {Santos}}, \bibinfo {author} {\bibfnamefont {V.}~\bibnamefont {Garz\'o}}, \
  and\ \bibinfo {author} {\bibfnamefont {J.~W.}\ \bibnamefont {Dufty}},\
  }\bibfield  {title} {\enquote {\bibinfo {title} {Inherent rheology of a
  granular fluid in uniform shear flow},}\ }\href@noop {} {\bibfield  {journal}
  {\bibinfo  {journal} {Phys. Rev. E}\ }\textbf {\bibinfo {volume} {69}},\
  \bibinfo {pages} {{061}{303}} (\bibinfo {year} {2004})}\BibitemShut {NoStop}%
\bibitem [{\citenamefont {Ikenberry}\ and\ \citenamefont
  {Truesdell}(1956)}]{IT56}%
  \BibitemOpen
  \bibfield  {author} {\bibinfo {author} {\bibfnamefont {E.}~\bibnamefont
  {Ikenberry}}\ and\ \bibinfo {author} {\bibfnamefont {C.}~\bibnamefont
  {Truesdell}},\ }\bibfield  {title} {\enquote {\bibinfo {title} {On the
  pressures and the flux of energy in a gas according to {M}axwell's kinetic
  theory},}\ }\href@noop {} {\bibfield  {journal} {\bibinfo  {journal} {J. Rat.
  Mech. Anal.}\ }\textbf {\bibinfo {volume} {5}},\ \bibinfo {pages} {1--54}
  (\bibinfo {year} {1956})}\BibitemShut {NoStop}%
\bibitem [{\citenamefont {McLennan}(1989)}]{M89}%
  \BibitemOpen
  \bibfield  {author} {\bibinfo {author} {\bibfnamefont {J.~A.}\ \bibnamefont
  {McLennan}},\ }\href@noop {} {\emph {\bibinfo {title} {Introduction to
  Nonequilibrium Statistical Mechanics}}}\ (\bibinfo  {publisher}
  {Prentic--Hall, New Yersey},\ \bibinfo {year} {1989})\BibitemShut {NoStop}%
\bibitem [{\citenamefont {Bird}(1994)}]{B94}%
  \BibitemOpen
  \bibfield  {author} {\bibinfo {author} {\bibfnamefont {G.~A.}\ \bibnamefont
  {Bird}},\ }\href@noop {} {\emph {\bibinfo {title} {Molecular Gas Dynamics and
  the Direct Simulation Monte Carlo of Gas Flows}}}\ (\bibinfo  {publisher}
  {Clarendon, Oxford},\ \bibinfo {year} {1994})\BibitemShut {NoStop}%
\bibitem [{\citenamefont {Brey}\ \emph {et~al.}(1999)\citenamefont {Brey},
  \citenamefont {Dufty},\ and\ \citenamefont {Santos}}]{BDS99}%
  \BibitemOpen
  \bibfield  {author} {\bibinfo {author} {\bibfnamefont {J.~J.}\ \bibnamefont
  {Brey}}, \bibinfo {author} {\bibfnamefont {J.~W.}\ \bibnamefont {Dufty}}, \
  and\ \bibinfo {author} {\bibfnamefont {A.}~\bibnamefont {Santos}},\
  }\bibfield  {title} {\enquote {\bibinfo {title} {Kinetic models for granular
  flow},}\ }\href@noop {} {\bibfield  {journal} {\bibinfo  {journal} {J. Stat.
  Phys.}\ }\textbf {\bibinfo {volume} {97}},\ \bibinfo {pages} {281--322}
  (\bibinfo {year} {1999})}\BibitemShut {NoStop}%
\bibitem [{\citenamefont {Sarracino}\ \emph {et~al.}(2010)\citenamefont
  {Sarracino}, \citenamefont {Villamaina}, \citenamefont {Costantini},\ and\
  \citenamefont {Puglisi}}]{SVCP10}%
  \BibitemOpen
  \bibfield  {author} {\bibinfo {author} {\bibfnamefont {A.}~\bibnamefont
  {Sarracino}}, \bibinfo {author} {\bibfnamefont {D.}~\bibnamefont
  {Villamaina}}, \bibinfo {author} {\bibfnamefont {G.}~\bibnamefont
  {Costantini}}, \ and\ \bibinfo {author} {\bibfnamefont {A.}~\bibnamefont
  {Puglisi}},\ }\bibfield  {title} {\enquote {\bibinfo {title} {Granular
  {B}rownian motion},}\ }\href@noop {} {\bibfield  {journal} {\bibinfo
  {journal} {J. Stat. Mech.}\ }\textbf {\bibinfo {volume} {P04013}} (\bibinfo
  {year} {2010})}\BibitemShut {NoStop}%
\bibitem [{\citenamefont {Bocquet}\ \emph
  {et~al.}(1994{\natexlab{a}})\citenamefont {Bocquet}, \citenamefont {Hansen},\
  and\ \citenamefont {Piasecki}}]{BHP94}%
  \BibitemOpen
  \bibfield  {author} {\bibinfo {author} {\bibfnamefont {L.}~\bibnamefont
  {Bocquet}}, \bibinfo {author} {\bibfnamefont {J.-P.}\ \bibnamefont {Hansen}},
  \ and\ \bibinfo {author} {\bibfnamefont {J.}~\bibnamefont {Piasecki}},\
  }\bibfield  {title} {\enquote {\bibinfo {title} {On the {B}rownian motion of
  a massive sphere suspended in a hard-sphere fluid. {II}. {M}olecular dynamics
  estimates of the friction coefficient},}\ }\href@noop {} {\bibfield
  {journal} {\bibinfo  {journal} {J. Stat. Phys.}\ }\textbf {\bibinfo {volume}
  {76}},\ \bibinfo {pages} {527--548} (\bibinfo {year}
  {1994}{\natexlab{a}})}\BibitemShut {NoStop}%
\bibitem [{\citenamefont {Bocquet}\ \emph
  {et~al.}(1994{\natexlab{b}})\citenamefont {Bocquet}, \citenamefont
  {Piasecki},\ and\ \citenamefont {Hansen}}]{BPH94}%
  \BibitemOpen
  \bibfield  {author} {\bibinfo {author} {\bibfnamefont {L.}~\bibnamefont
  {Bocquet}}, \bibinfo {author} {\bibfnamefont {J.}~\bibnamefont {Piasecki}}, \
  and\ \bibinfo {author} {\bibfnamefont {J.-P.}\ \bibnamefont {Hansen}},\
  }\bibfield  {title} {\enquote {\bibinfo {title} {On the {B}rownian motion of
  a massive sphere suspended in a hard-sphere fluid. {I}. {M}ultiple-time-scale
  analysis and microscopic expression of the friction coefficient},}\
  }\href@noop {} {\bibfield  {journal} {\bibinfo  {journal} {J. Stat. Phys.}\
  }\textbf {\bibinfo {volume} {76}},\ \bibinfo {pages} {505--526} (\bibinfo
  {year} {1994}{\natexlab{b}})}\BibitemShut {NoStop}%
\bibitem [{\citenamefont {Santos}(2003)}]{S03a}%
  \BibitemOpen
  \bibfield  {author} {\bibinfo {author} {\bibfnamefont {A.}~\bibnamefont
  {Santos}},\ }\bibfield  {title} {\enquote {\bibinfo {title} {Granular fluid
  thermostated by a bath of elastic hard spheres},}\ }\href@noop {} {\bibfield
  {journal} {\bibinfo  {journal} {Phys. Rev. E}\ }\textbf {\bibinfo {volume}
  {67}},\ \bibinfo {pages} {051101} (\bibinfo {year} {2003})}\BibitemShut
  {NoStop}%
\bibitem [{\citenamefont {Rodr\'iguez}\ \emph {et~al.}(1983)\citenamefont
  {Rodr\'iguez}, \citenamefont {Salinas-Rodr\'iguez},\ and\ \citenamefont
  {Dufty}}]{RSD83}%
  \BibitemOpen
  \bibfield  {author} {\bibinfo {author} {\bibfnamefont {R.~F.}\ \bibnamefont
  {Rodr\'iguez}}, \bibinfo {author} {\bibfnamefont {E.}~\bibnamefont
  {Salinas-Rodr\'iguez}}, \ and\ \bibinfo {author} {\bibfnamefont {J.~W.}\
  \bibnamefont {Dufty}},\ }\bibfield  {title} {\enquote {\bibinfo {title}
  {Fokker-{P}lanck and {L}angevin descriptions of fluctuations in uniform shear
  flow},}\ }\href@noop {} {\bibfield  {journal} {\bibinfo  {journal} {J. Stat.
  Phys.}\ }\textbf {\bibinfo {volume} {32}},\ \bibinfo {pages} {279--298}
  (\bibinfo {year} {1983})}\BibitemShut {NoStop}%
\bibitem [{\citenamefont {G\'omez~Gonz\'alez}\ and\ \citenamefont
  {Garz\'o}(2020)}]{GG20}%
  \BibitemOpen
  \bibfield  {author} {\bibinfo {author} {\bibfnamefont {R.}~\bibnamefont
  {G\'omez~Gonz\'alez}}\ and\ \bibinfo {author} {\bibfnamefont
  {V.}~\bibnamefont {Garz\'o}},\ }\bibfield  {title} {\enquote {\bibinfo
  {title} {Non-{N}ewtonian rheology in inertial suspensions of inelastic rough
  hard spheres under simple shear flow},}\ }\href@noop {} {\bibfield  {journal}
  {\bibinfo  {journal} {Phys. Fluids}\ }\textbf {\bibinfo {volume} {32}},\
  \bibinfo {pages} {073315} (\bibinfo {year} {2020})}\BibitemShut {NoStop}%
\bibitem [{\citenamefont {Santos}\ \emph {et~al.}(1998)\citenamefont {Santos},
  \citenamefont {Montanero}, \citenamefont {Dufty},\ and\ \citenamefont
  {Brey}}]{SMDB98}%
  \BibitemOpen
  \bibfield  {author} {\bibinfo {author} {\bibfnamefont {A.}~\bibnamefont
  {Santos}}, \bibinfo {author} {\bibfnamefont {J.~M.}\ \bibnamefont
  {Montanero}}, \bibinfo {author} {\bibfnamefont {J.~W.}\ \bibnamefont
  {Dufty}}, \ and\ \bibinfo {author} {\bibfnamefont {J.~J.}\ \bibnamefont
  {Brey}},\ }\bibfield  {title} {\enquote {\bibinfo {title} {Kinetic model for
  the hard-sphere fluid and solid},}\ }\href@noop {} {\bibfield  {journal}
  {\bibinfo  {journal} {Phys. Rev. E}\ }\textbf {\bibinfo {volume} {57}},\
  \bibinfo {pages} {1644--1660} (\bibinfo {year} {1998})}\BibitemShut {NoStop}%
\bibitem [{\citenamefont {G\'omez~Gonz\'alez}\ \emph
  {et~al.}(2023)\citenamefont {G\'omez~Gonz\'alez}, \citenamefont {Abad},
  \citenamefont {Bravo~Yuste},\ and\ \citenamefont {Garz\'o}}]{GASG23}%
  \BibitemOpen
  \bibfield  {author} {\bibinfo {author} {\bibfnamefont {R.}~\bibnamefont
  {G\'omez~Gonz\'alez}}, \bibinfo {author} {\bibfnamefont {E.}~\bibnamefont
  {Abad}}, \bibinfo {author} {\bibfnamefont {S.}~\bibnamefont {Bravo~Yuste}}, \
  and\ \bibinfo {author} {\bibfnamefont {V.}~\bibnamefont {Garz\'o}},\
  }\bibfield  {title} {\enquote {\bibinfo {title} {Diffusion of intruders in
  granular suspensions: Enskog theory and random walk interpretation},}\
  }\href@noop {} {\bibfield  {journal} {\bibinfo  {journal} {Phys. Rev. E}\
  }\textbf {\bibinfo {volume} {108}},\ \bibinfo {pages} {024903} (\bibinfo
  {year} {2023})}\BibitemShut {NoStop}%
\bibitem [{\citenamefont {Koch}\ and\ \citenamefont {Hill}(2001)}]{KH01}%
  \BibitemOpen
  \bibfield  {author} {\bibinfo {author} {\bibfnamefont {D.~L.}\ \bibnamefont
  {Koch}}\ and\ \bibinfo {author} {\bibfnamefont {R.~J.}\ \bibnamefont
  {Hill}},\ }\bibfield  {title} {\enquote {\bibinfo {title} {Inertial effects
  in suspensions and porous-media flows},}\ }\href@noop {} {\bibfield
  {journal} {\bibinfo  {journal} {Annu. Rev. Fluid Mech.}\ }\textbf {\bibinfo
  {volume} {33}},\ \bibinfo {pages} {619--647} (\bibinfo {year}
  {2001})}\BibitemShut {NoStop}%
\bibitem [{\citenamefont {Tenneti}\ and\ \citenamefont
  {Subramaniam}(2014)}]{TS14}%
  \BibitemOpen
  \bibfield  {author} {\bibinfo {author} {\bibfnamefont {S.}~\bibnamefont
  {Tenneti}}\ and\ \bibinfo {author} {\bibfnamefont {S.}~\bibnamefont
  {Subramaniam}},\ }\bibfield  {title} {\enquote {\bibinfo {title}
  {Particle-resolved direct numerical simulation fo gas-solid flow model
  development},}\ }\href@noop {} {\bibfield  {journal} {\bibinfo  {journal}
  {Annu. Rev. Fluid Mech.}\ }\textbf {\bibinfo {volume} {46}},\ \bibinfo
  {pages} {199--230} (\bibinfo {year} {2014})}\BibitemShut {NoStop}%
\bibitem [{\citenamefont {Wang}(2020)}]{W20}%
  \BibitemOpen
  \bibfield  {author} {\bibinfo {author} {\bibfnamefont {J.}~\bibnamefont
  {Wang}},\ }\bibfield  {title} {\enquote {\bibinfo {title} {Continuum theory
  for dense gas-solid flow: A state-of-the-art review},}\ }\href@noop {}
  {\bibfield  {journal} {\bibinfo  {journal} {Chem. Eng. Sci.}\ }\textbf
  {\bibinfo {volume} {215}},\ \bibinfo {pages} {115428} (\bibinfo {year}
  {2020})}\BibitemShut {NoStop}%
\bibitem [{\citenamefont {Liu}\ \emph {et~al.}(2020)\citenamefont {Liu},
  \citenamefont {Ge},\ and\ \citenamefont {Wang}}]{LGW20}%
  \BibitemOpen
  \bibfield  {author} {\bibinfo {author} {\bibfnamefont {X.}~\bibnamefont
  {Liu}}, \bibinfo {author} {\bibfnamefont {W.}~\bibnamefont {Ge}}, \ and\
  \bibinfo {author} {\bibfnamefont {L.}~\bibnamefont {Wang}},\ }\bibfield
  {title} {\enquote {\bibinfo {title} {Scale and structure dependent drag in
  gas–solid flows},}\ }\href@noop {} {\bibfield  {journal} {\bibinfo
  {journal} {AIChE}\ }\textbf {\bibinfo {volume} {66}},\ \bibinfo {pages}
  {e16883} (\bibinfo {year} {2020})}\BibitemShut {NoStop}%
\bibitem [{\citenamefont {Batchelor}(1972)}]{B72}%
  \BibitemOpen
  \bibfield  {author} {\bibinfo {author} {\bibfnamefont {G.~K.}\ \bibnamefont
  {Batchelor}},\ }\bibfield  {title} {\enquote {\bibinfo {title} {Sedimentation
  in a dilute dispersion of spheres},}\ }\href@noop {} {\bibfield  {journal}
  {\bibinfo  {journal} {J. Fluid Mech.}\ }\textbf {\bibinfo {volume} {52}},\
  \bibinfo {pages} {245--268} (\bibinfo {year} {1972})}\BibitemShut {NoStop}%
\bibitem [{\citenamefont {Brady}\ and\ \citenamefont {Bossis}(1988)}]{BB88}%
  \BibitemOpen
  \bibfield  {author} {\bibinfo {author} {\bibfnamefont {J.~F.}\ \bibnamefont
  {Brady}}\ and\ \bibinfo {author} {\bibfnamefont {G}~\bibnamefont {Bossis}},\
  }\bibfield  {title} {\enquote {\bibinfo {title} {Stokesian dynamics},}\
  }\href@noop {} {\bibfield  {journal} {\bibinfo  {journal} {Ann. Rev. Fluid
  Mech.}\ }\textbf {\bibinfo {volume} {20}},\ \bibinfo {pages} {111--157}
  (\bibinfo {year} {1988})}\BibitemShut {NoStop}%
\bibitem [{\citenamefont {Louge}\ \emph {et~al.}(1991)\citenamefont {Louge},
  \citenamefont {Mastorakos},\ and\ \citenamefont {Jenkins}}]{LMJ91}%
  \BibitemOpen
  \bibfield  {author} {\bibinfo {author} {\bibfnamefont {M.}~\bibnamefont
  {Louge}}, \bibinfo {author} {\bibfnamefont {E.}~\bibnamefont {Mastorakos}}, \
  and\ \bibinfo {author} {\bibfnamefont {J.~T.}\ \bibnamefont {Jenkins}},\
  }\bibfield  {title} {\enquote {\bibinfo {title} {The role of particle
  collisions in pneumatic transport},}\ }\href@noop {} {\bibfield  {journal}
  {\bibinfo  {journal} {J. Fluid Mech.}\ }\textbf {\bibinfo {volume} {231}},\
  \bibinfo {pages} {345--359} (\bibinfo {year} {1991})}\BibitemShut {NoStop}%
\bibitem [{\citenamefont {Wylie}\ \emph
  {et~al.}(2009{\natexlab{b}})\citenamefont {Wylie}, \citenamefont {Zhang},
  \citenamefont {Li},\ and\ \citenamefont {Hengyi}}]{WZL09}%
  \BibitemOpen
  \bibfield  {author} {\bibinfo {author} {\bibfnamefont {J.~J.}\ \bibnamefont
  {Wylie}}, \bibinfo {author} {\bibfnamefont {Q.}~\bibnamefont {Zhang}},
  \bibinfo {author} {\bibfnamefont {Yun}\ \bibnamefont {Li}}, \ and\ \bibinfo
  {author} {\bibfnamefont {Xu}~\bibnamefont {Hengyi}},\ }\bibfield  {title}
  {\enquote {\bibinfo {title} {Driven inelastic-particle systems with drag},}\
  }\href@noop {} {\bibfield  {journal} {\bibinfo  {journal} {Phys. Rev. E}\
  }\textbf {\bibinfo {volume} {79}},\ \bibinfo {pages} {031301} (\bibinfo
  {year} {2009}{\natexlab{b}})}\BibitemShut {NoStop}%
\bibitem [{\citenamefont {Parmentier}\ and\ \citenamefont
  {Simonin}(2012)}]{PS12}%
  \BibitemOpen
  \bibfield  {author} {\bibinfo {author} {\bibfnamefont {J.-F.}\ \bibnamefont
  {Parmentier}}\ and\ \bibinfo {author} {\bibfnamefont {O.}~\bibnamefont
  {Simonin}},\ }\bibfield  {title} {\enquote {\bibinfo {title} {Transition
  models from the quenched to ignited states for flows of inertial particles
  suspended in a simple sheared viscous fluid},}\ }\href@noop {} {\bibfield
  {journal} {\bibinfo  {journal} {J. Fluid Mech.}\ }\textbf {\bibinfo {volume}
  {711}},\ \bibinfo {pages} {147--160} (\bibinfo {year} {2012})}\BibitemShut
  {NoStop}%
\bibitem [{\citenamefont {Wang}\ \emph {et~al.}(2014)\citenamefont {Wang},
  \citenamefont {Grob}, \citenamefont {Zippelius},\ and\ \citenamefont
  {Sperl}}]{WGZS14}%
  \BibitemOpen
  \bibfield  {author} {\bibinfo {author} {\bibfnamefont {T.}~\bibnamefont
  {Wang}}, \bibinfo {author} {\bibfnamefont {M.}~\bibnamefont {Grob}}, \bibinfo
  {author} {\bibfnamefont {A.}~\bibnamefont {Zippelius}}, \ and\ \bibinfo
  {author} {\bibfnamefont {M.}~\bibnamefont {Sperl}},\ }\bibfield  {title}
  {\enquote {\bibinfo {title} {Active microrhelogy of driven granular
  particles},}\ }\href@noop {} {\bibfield  {journal} {\bibinfo  {journal}
  {Phys. Rev. E}\ }\textbf {\bibinfo {volume} {89}},\ \bibinfo {pages}
  {{042}{209}} (\bibinfo {year} {2014})}\BibitemShut {NoStop}%
\bibitem [{\citenamefont {Puglisi}\ \emph {et~al.}(1998)\citenamefont
  {Puglisi}, \citenamefont {Loreto}, \citenamefont {Marconi}, \citenamefont
  {Petri},\ and\ \citenamefont {Vulpiani}}]{PLMPV98}%
  \BibitemOpen
  \bibfield  {author} {\bibinfo {author} {\bibfnamefont {A.}~\bibnamefont
  {Puglisi}}, \bibinfo {author} {\bibfnamefont {V.}~\bibnamefont {Loreto}},
  \bibinfo {author} {\bibfnamefont {U.~M.~B.}\ \bibnamefont {Marconi}},
  \bibinfo {author} {\bibfnamefont {A.}~\bibnamefont {Petri}}, \ and\ \bibinfo
  {author} {\bibfnamefont {A.}~\bibnamefont {Vulpiani}},\ }\bibfield  {title}
  {\enquote {\bibinfo {title} {Clustering and non-{G}aussian behavior in
  granular matter},}\ }\href@noop {} {\bibfield  {journal} {\bibinfo  {journal}
  {Phys. Rev. Lett.}\ }\textbf {\bibinfo {volume} {81}},\ \bibinfo {pages}
  {3848--3851} (\bibinfo {year} {1998})}\BibitemShut {NoStop}%
\bibitem [{\citenamefont {Brown}\ and\ \citenamefont {Jeager}(2014)}]{BJ14}%
  \BibitemOpen
  \bibfield  {author} {\bibinfo {author} {\bibfnamefont {E.}~\bibnamefont
  {Brown}}\ and\ \bibinfo {author} {\bibfnamefont {H.~M.}\ \bibnamefont
  {Jeager}},\ }\bibfield  {title} {\enquote {\bibinfo {title} {Dynamic jamming
  point for shear thickening suspensions},}\ }\href@noop {} {\bibfield
  {journal} {\bibinfo  {journal} {Rep. Prog. Phys.}\ }\textbf {\bibinfo
  {volume} {77}},\ \bibinfo {pages} {046602} (\bibinfo {year}
  {2014})}\BibitemShut {NoStop}%
\bibitem [{\citenamefont {Bourrianne}\ \emph {et~al.}(2022)\citenamefont
  {Bourrianne}, \citenamefont {Niggel}, \citenamefont {Polly}, \citenamefont
  {Divoux},\ and\ \citenamefont {McKinley}}]{BNPDM22}%
  \BibitemOpen
  \bibfield  {author} {\bibinfo {author} {\bibfnamefont {P.}~\bibnamefont
  {Bourrianne}}, \bibinfo {author} {\bibfnamefont {V.}~\bibnamefont {Niggel}},
  \bibinfo {author} {\bibfnamefont {G.}~\bibnamefont {Polly}}, \bibinfo
  {author} {\bibfnamefont {T.}~\bibnamefont {Divoux}}, \ and\ \bibinfo {author}
  {\bibfnamefont {G.~H.}\ \bibnamefont {McKinley}},\ }\bibfield  {title}
  {\enquote {\bibinfo {title} {Tuning the shear thickening of suspensions
  through surface roughness and physico-chemical interactions},}\ }\href@noop
  {} {\bibfield  {journal} {\bibinfo  {journal} {Phys. Rev. Res.}\ }\textbf
  {\bibinfo {volume} {4}},\ \bibinfo {pages} {033062} (\bibinfo {year}
  {2022})}\BibitemShut {NoStop}%
\end{thebibliography}%
\end{document}